\documentclass[aps,prl,superscriptaddress,twocolumn]{revtex4-1}
\usepackage[utf8]{inputenc}
\usepackage{tikz-cd}
\usepackage{color}
\usepackage{dsfont}
\usepackage{bm}
\usepackage{mathtools}
\usepackage{amsthm}
\usepackage{amsfonts,amssymb,amsmath}
\usepackage[hidelinks]{hyperref}
\usepackage{physics} \usepackage{multirow}
\usepackage{graphicx}

\usepackage[export]{adjustbox}
\usepackage{verbatim}
\usepackage[normalem]{ulem}

\begin{document}

%\title{Phase transition in the projected ensemble from random permutation quantum dynamics}
\title{Coherence-induced deep thermalization transition in  random permutation quantum dynamics}
%\title{Deep thermalization phase transition in random permutation quantum dynamics}
\author{Chang Liu}
\affiliation{Department of Physics, National University of Singapore, Singapore 117551}
\author{Matteo Ippoliti}
\affiliation{Department of Physics, The University of Texas at Austin, Austin, TX 78712, USA}
\author{Wen Wei Ho}
\altaffiliation{\href{mailto:wenweiho@nus.edu.sg}{wenweiho@nus.edu.sg}}
\affiliation{Department of Physics, National University of Singapore, Singapore 117551}
\affiliation{Centre for Quantum Technologies, National University of Singapore,   Singapore 117543}

\begin{abstract} 
We report a phase transition in the projected ensemble --- the collection of post-measurement wavefunctions of a local subsystem obtained by measuring its complement. 
The transition emerges in systems undergoing random permutation dynamics, a type of quantum time evolution wherein computational basis states are shuffled without creating superpositions. It separates a phase exhibiting  deep thermalization, where the projected ensemble is distributed over Hilbert space in a maximally entropic fashion (Haar-random), from a phase where it is minimally entropic (``classical bit-string ensemble''). 
Crucially, this deep thermalization transition is invisible to the subsystem's density matrix, which always exhibits thermalization to infinite-temperature across the phase diagram. 
Through a combination of analytical arguments and  numerical simulations, we show that the transition is tuned by the total amount of {\it coherence} injected by the input state and the measurement basis, and is exhibited robustly across different microscopic models. 
Our findings represent a novel form of ergodicity-breaking universality in quantum many-body dynamics, characterized not by a failure of regular thermalization, but rather by a failure of deep thermalization. 
\end{abstract}
\date{\today}
\maketitle

{\it Introduction.}---Understanding universal behaviors of complex quantum systems out of equilibrium is a central goal of modern physics, with implications in statistical mechanics~\cite{rigol2008thermalization, Eisert2015}, condensed matter~\cite{ Nandkishore_review2015, RevModPhys.91.021001, RevModPhys.97.025004}, high energy physics~\cite{Yasuhiro_Sekino_2008, Shenker2014, Maldacena2016}, and quantum information science~\cite{Patrick_Hayden_2007, annurev:/content/journals/10.1146/annurev-conmatphys-031720-030658}.  
Recently, a new universal feature was discovered in the dynamics of quantum many-body systems: the collection of conditional states of a local subsystem obtained by measuring its environment, known as the {\it projected ensemble}~(PE)~\cite{choi2023preparing, cotler2023emergent, ho2022exact}, was typically found to  approach universal distributions at late times in dynamics, which satisfy generalized maximum-entropy principles~\cite{mark2024maximum, liu2024deep, chang2025deep}.
For example,  
in spin systems, one obtains the Haar ensemble (the uniform distribution over the Hilbert space)~\cite{cotler2023emergent, ho2022exact,ippoliti2022solvable, ippoliti2023dynamical} or the Scrooge ensemble (a deformation thereof) if conservation laws are present~\cite{cotler2023emergent,mark2024maximum,mcginley2025scrooge,mok2026nature}; analogous maximally-entropic distributions arise also for systems of Gaussian fermions and bosons~\cite{lucas2023generalized,bejan2025matchgate,liu2024deep}. 
The emergence of such universal ergodic ensembles has been dubbed {\it deep thermalization}~\cite{choi2023preparing,cotler2023emergent,ho2022exact,claeys2022emergent,
ippoliti2022solvable,ippoliti2023dynamical,
lucas2023generalized,bhore2023deep,chan2024projected,varikuti2024unraveling, mark2024maximum,lucas2023generalized,bejan2025matchgate,liu2024deep,shrotriya2025nonlocality,chang2025deep,zhang2025holographic}, as it constitutes a stronger, more fine-grained notion of quantum equilibration going beyond ``regular'' thermalization of the reduced density matrix to the Gibbs state. 
The inception of deep thermalization has led to a flurry of research activity, including studies on the timescales of convergence~\cite{ippoliti2022solvable,ippoliti2023dynamical}, generalizations to open quantum systems~\cite{yu2025mixed,sherry2025mixed}, connections to computational complexity and cryptography~\cite{chakraborty2025fast,zhang2025holographic}, as well as experimental realizations~\cite{choi2023preparing}. 

\begin{figure}[t]
\includegraphics[width=0.5\textwidth]{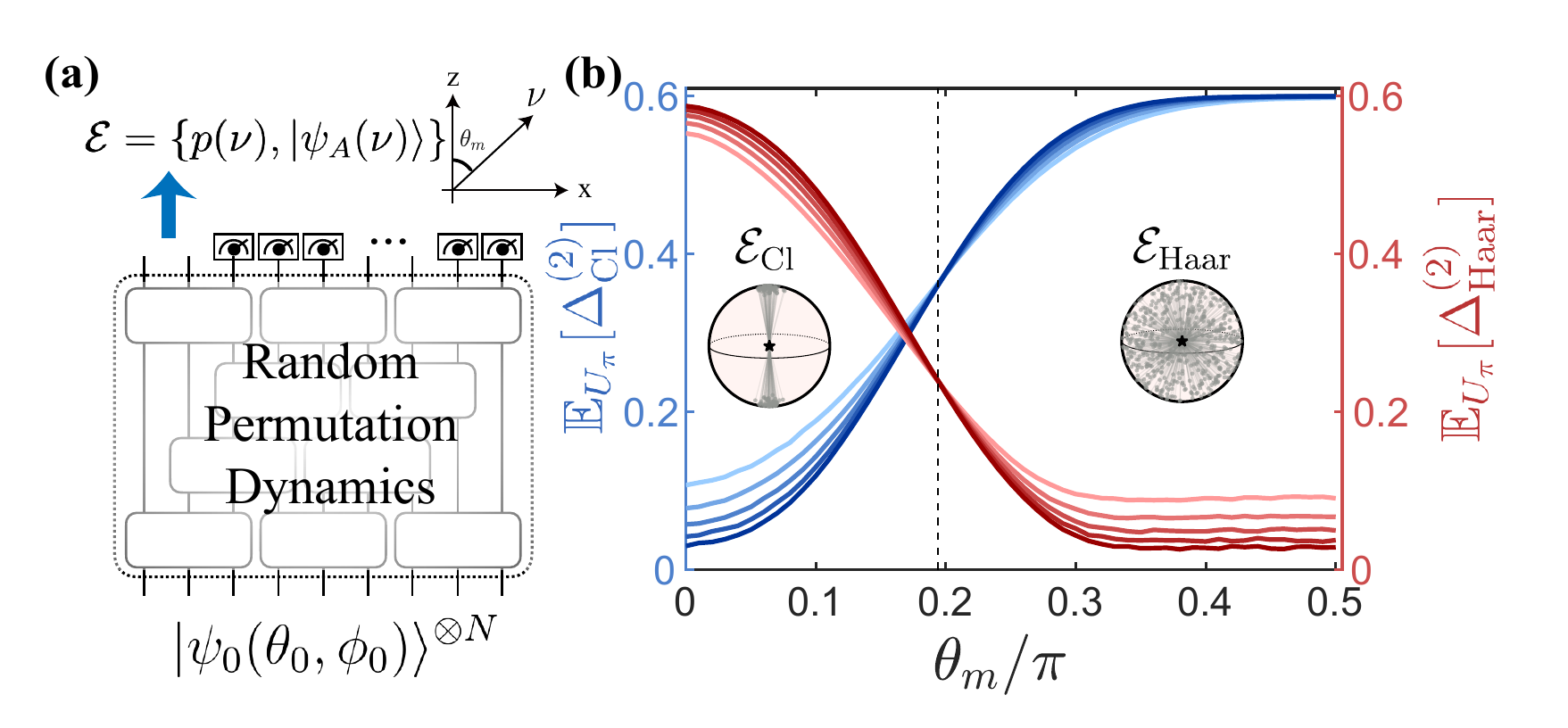}
\caption{(a) Projected ensemble~(PE) formed under random permutation dynamics. A single global random permutation unitary~(dotted box) models the behavior of a deep quantum circuit made of local random permutation gates~(brickwork circuit). For the tilted-basis model, input states and measurement basis are uniform product states, specified by Bloch angles $(\theta_0$\,$,$\,$\phi_0)$ and $(\theta_m$\,$,$\,$\phi_m)$ respectively. 
(b) $k$\,$=$\,$2$ trace distances of the PE (for $N_A$\,$=$\,$2$) from the classical bit-string ensemble $\mathcal{E}_\text{Cl}$ and Haar ensemble $\mathcal{E}_\text{Haar}$, generated from  the tilted-basis model  with $\theta_0$\,$=$\,$\phi_0$\,$=$\,$\pi/4$, $\phi_m$\,$=$\,$0$, and variable $\theta_m$. Different intensities indicate  different system sizes $N$\,$=$\,$16,18,20,22,24$~(lighter to darker). 
There is a common crossing at $\theta_m^*$\,$\approx$\,$0.193\pi$ 
for both distances across all system sizes, signaling a singular change of the limiting PE.
%; \matteo{this last part can be dropped if we need to save space (it's addressed in main text)}
% on either side of this point, the distances flow toward zero or to their maximal value, indicating a singular change of the limiting projected ensemble.
}
\label{Fig:1}
\end{figure}

In this Letter, we uncover a striking exception to the aforementioned maximum-entropy paradigm within a class of quantum dynamics. Namely, we report a sharp transition in the PE separating a phase that exhibits deep thermalization from a phase that robustly fails to do so. 
The transition is completely invisible to the reduced density matrix: both phases always appear thermal at infinite temperature. Our findings  thus represent a novel form of  ``deep'' ergodicity-breaking in quantum many-body systems, detectable not at the level of local  expectation values, but rather in higher-moment observables.
 
Concretely, we study {\it random permutation dynamics} (RPD)~[Fig.~\ref{Fig:1}(a)], a type of quantum time-evolution where computational basis states are randomly permuted~\cite{Aldana_2011, bertini2025permutation, szasz2025entanglement,bertini2025permutation_chaotic}. The transition is tuned by the amount of {\it coherence} (the information-theoretic resource of superposition~\cite{baumgratz2014quantifying, PhysRevLett.116.120404, RevModPhys.89.041003}) in the system, which is preserved under RPD and thus depends only on the choice of input state and final measurement basis. 
The high- and low-coherence phases yield, respectively, the Haar ensemble $\mathcal{E}_\text{Haar}$, wherein states are uniformly  distributed in Hilbert space, and the ``classical bit-string ensemble'' $\mathcal{E}_\text{Cl}$, wherein states are distributed uniformly but only over the computational basis~[Fig.~\ref{Fig:1}(b)]; these are respectively the maximum- and minimum-entropy PEs compatible with an infinite-temperature density matrix.
We establish the universality of this phenomenon across two different microscopic models theoretically and numerically; 
in particular, one of the models allows an analytical determination of the phase boundary. 
More generally, we argue the transition occurs generically for typical scrambling quantum dynamics, as long as it is coherence-preserving. 
Lastly, we discuss generalizations of our analysis to transitions of other resources within the PE, like imaginarity~\cite{hickey2018quantifying,wu2021resource}, magic~\cite{PhysRevA.86.052329, PhysRevLett.118.090501, PRXQuantum.3.020333} or non-Gaussianity~\cite{PhysRevA.97.052317, PhysRevA.98.022335},  suggesting the possibility of yet more novel universality classes in quantum many-body dynamics.  
 
{\it Projected ensemble from random permutation dynamics.}---Consider an $N$-qubit system bipartitioned into $A$ and $B$ (comprising $N_A$ and $N_B$ qubits respectively) with initial state $|\Psi_0\rangle$. This state is acted upon by a randomly-chosen global permutation unitary $U_\pi$, to yield a final state $|\Psi_\pi\rangle$\,$=$\,$U_\pi |\Psi_0\rangle$, which we denote as a {\it random permutation state}~(RPS). Specifically, $U_\pi$ is a unitary which permutes computational basis states $|z\rangle$ (with $z$\,$\in$\,$\{0,1\}^N$) for some uniformly randomly chosen permutation $\pi$ from the symmetric group $S_{2^N}$, i.e.,~$U_\pi|z\rangle$\,$=$\,$|\pi(z)\rangle$.
The  application of a single global permutation unitary can be understood as modeling the late-time behavior of  quantum circuits composed of local permutation unitaries~[see e.g.~Fig.~\ref{Fig:1}(a)], and indeed in the End Matter, we show that such a simplification does not affect the physics to be discussed. 
RPD,  long studied in the classical literature~\cite{gowers1996almost,hoory2005simplepermutations,brodsky2008simplepermutations}, has recently found applications in the quantum setting. This is because despite its ``classical'' action on bit-string states, it can nevertheless reproduce general aspects of quantum dynamics like  growth of  state and operator entanglement~\cite{Gopalakrishnan2018automata,iaconis_subdiffusion_2019,iaconis2020measurement,bertini2025permutation,szasz2025entanglement}, %\matteo{sarang's is non-random but it seems relevant nonetheless - can remove if you prefer}
decay of out-of-time order correlators~\cite{iaconis_automata_2021,bertini2025permutation_chaotic}, and formation of state designs~\cite{Feng2025,lee2024fastpseudothermalization} when evolving from generic initial states.

Our aim is to characterize the limiting form of the PE of a fixed local subsystem $A$ generated from a typical RPS given a fixed choice of initial state $|\Psi_0\rangle$, and measurements of the complementary region $B$ in various bases. Here $B$ is interpreted as the `bath', and assumed much larger than $A$. 
We consider two models: (i)~the ``tilted-basis model'', with initial states  taken as uniform product states $|\Psi_0\rangle$\,$=$\,$(\cos (\theta_0/2)|0\rangle + e^{i\phi_0}\sin(\theta_0/2)|1\rangle)^{\otimes N}$  
and measurements also along a uniform local basis but with direction $\hat{n}$\,$=$\,$(\sin\theta_m\cos\phi_m$\,$,$\,$\sin\theta_m\sin\phi_m$\,$,$\,$\cos\theta_m)$. We henceforth set $\phi_m$\,$=$\,$0$ for simplicity; $\theta_m$ thus continuously tunes  the measurement basis from the $z$- to the $x$-axis.
(ii)~The ``mixed-basis model", with initial states  $|\Psi_0\rangle$\,$:=$\,$\ket{0}^{\otimes (1-\alpha_0)N} \otimes \ket{\mathrm{Y}_+}^{\otimes \alpha_0 N}$ and $|\mathrm{Y}_+\rangle$\,$=$\,$\frac{1}{\sqrt{2}}(|0\rangle + i |1\rangle)$, and measurement schemes in which $(1-\alpha_m)N_B$ qubits ($\alpha_m N_B$ qubits) are measured in the $z$-basis~($x$-basis). 
Intensive parameters $\alpha_0$\,$\in$\,$\{0,\cdots,N-1\}/N, \alpha_m$\,$\in$\,$\{0,\cdots,N_B-1\}/N_B$ tune the amount of superposition (over $|z\rangle$) of the initial state and measurement basis respectively, which play analogous roles as $\theta_0$\,$,$\,$\theta_m$; thus, the mixed-basis model can be thought of as a discrete analog of the tilted-basis model. 

Upon measuring, one obtains a  bit-string   $\nu$\,$\in$\,$\{0,1\}^{N_B}$ with Born probability $p(\nu)$, together with a post-measurement pure state  $|\psi_A(\nu)\rangle$\,$=$\,$(I_A$\,$\otimes $\,$\langle\Phi_\nu|_B)|\Psi_\pi\rangle/\sqrt{p(\nu)}$ on subsystem $A$. 
Each $0(1)$ bit in $\nu$ denotes a measurement outcome aligned with(against) 
the particular local measurement basis, while $|\Phi_\nu\rangle_B$ is the product state on $B$ associated with measurement outcome $\nu$.  The PE is defined as the ensemble of such projected states with probabilities
\begin{equation}
\mathcal{E}_\text{PE}
:= \bigl\{p(\nu),\, |\psi_A(\nu)\rangle \bigr\}. 
\end{equation}
Importantly, while the PE completely specifies the reduced density matrix (RDM)~$\rho_A=\mathbb{E}_{\psi_A \sim \mathcal{E}_\text{PE}}[|\psi_A\rangle\langle \psi_A|]$, it contains strictly more information, as it describes a {\it distribution} over the Hilbert space $\mathcal{H}_A$ of $A$. 
Deep thermalization is  the emergence of universal ergodic distributions describing the PE in the thermodynamic limit ($N_B$\,$\to$\,$\infty$, fixing $N_A$) consistent with generalized maximum entropy principles. 
A standard way to quantify the convergence of the PE $\mathcal{E}_\text{PE}$ to a target distribution $\mathcal{E}_*$ is to compare the closeness of their $k$-th moments in the thermodynamic limit, such as through the trace distance $\Delta^{(k)}_*$\,$:=$\,$\frac{1}{2}\| \rho^{(k)}_\text{PE} - \rho^{(k)}_*\|_1$, where $\rho^{(k)}_{(\cdot)}$\,$:=$\,$\mathbb{E}_{\psi_A \sim \mathcal{E}_{(\cdot)}}[|\psi_A\rangle\langle \psi_A|^{\otimes k}]$~\cite{cotler2023emergent}. 

{\it Deep thermalization and deep ergodicity-breaking in the tilted-basis model.}--- We begin by analyzing the first moment of the PE within the tilted-basis model. This is the RDM $\rho_A$, which is independent of the   measurement basis direction on $B$.
We can rigorously show that local RDMs of almost all RPSs in the tilted-basis model are close to maximally-mixed:\\
{\bf Theorem 1.} {\it 
Let $|\Psi_\pi\rangle$ be an $N$-qubit RPS in the tilted-basis model specified by an initial state $|\Psi_0\rangle$ with Bloch-sphere angles $(\theta_0,\phi_0)$ not in $\{\theta_0$\,$=$\,$0,\theta_0$\,$=$\,$\pi,(\theta_0$\,$=$\,$\tfrac{\pi}{2},\phi_0$\,$=$\,$0)\}$. For any $\epsilon$\,$>$\,$0$, 
\begin{align}
\mathbb{P}\left(\left\|\rho_A - \frac{\mathbb{I}_A}{d_A}\right\|_1 > \epsilon\right) 
< \frac{(d^{-\alpha_0}+ d^{-2\beta_0})C+ d^{-1}D}{\epsilon^2},
\end{align}
where 
$d_A$\,$=$\,$2^{N_A},d$\,$=$\,$2^N$ are the Hilbert-space dimensions of $A$ and the full system respectively, 
$0$\,$<$\,$\alpha_0, \beta_0$\,$<$\,$1$ are factors determined by $(\theta_0,\phi_0)$, 
and $C$\,$,$\,$D$ are $O(1)$ in system-size factors (see the Supplemental Material~(SM)~\cite{supp} for details). 
}

The proof~\cite{supp} involves Weingarten calculus on the symmetric group~\cite{collins2025weingarten} and a Markov inequality to bound fluctuations from the maximally-mixed state.

Theorem 1 establishes that locally, the system almost always exhibits regular thermalization to the featureless infinite temperature state. 
We next turn to the higher moments $(k\geq2)$, which can now depend nontrivially on the choice of measurement basis. As a starting prediction, we employ the 
%\suggestion{``equivalence class''}
{version of the}
maximum-entropy principle~(MEP) put forth by~\cite{chang2025deep}. 
It  firstly prescribes that the exponentially many measurement outcomes $\nu$ entering the PE can be grouped into  $O(N_B)$ equivalence classes $[\nu]$ defined by their Hamming weight (number of $1$-bits)~\footnote{This is because the class of generator states (RPSs in this case) is invariant under qubit swaps in $B$.}; each class yields a density matrix $\rho_{[\nu]}$ on average over RPSs. Then, it posits that the PE for a typical RPS is a statistical mixture across equivalence classes of the pure-state unraveling of $\rho_{[\nu]}$ with least {\it accessible information}~\cite{jozsa1994lower} (the maximum amount of classical information extractable from the quantum-state ensemble); this yields the so-called ``generalized Scrooge ensemble''~\cite{mark2024maximum,chang2025deep}. In the SM~\cite{supp}, we present  details of the analyses following the principle. We find that independent of measurement direction $\hat{n}$, the generalized Scrooge ensemble is the  Haar ensemble 
\begin{align}
\mathcal{E}_\text{Haar} := \{d\psi_A, |\psi_A\rangle\},
\end{align}
 with $d\psi_A$ the Haar measure on $\mathcal{H}_A$. Thus, according to the MEP, we should expect that locally, the system is  featureless not only at the level of the density matrix, but also of the PE as a whole. 
Indeed, when measurements are along the $x$-basis~($\theta_m$\,$=$\,$\pi/2$), this prediction can be tested through an explicit computation of the expected PE:  $\mathbb{E}_{U_\pi}[\mathcal{E}_\text{PE}]$,$=$\,$\mathcal{E}_\text{Haar}$ (averaged over RPS for fixed $|\Psi_0\rangle$ with $\theta_0,\phi_0$\,$\notin$\,$\{0,\pi\}$), using Weingarten calculus and combinatorics of set partitions; see SM~\cite{supp} for details.  While this does not show convergence of $\mathcal{E}_\text{PE}$ to $\mathcal{E}_\text{Haar}$ for {\it individual} RPSs, it is a nontrivial necessary condition.

However, the MEP is only a guiding principle, and it is possible that the assumptions underlying its applicability fail, such that the limiting PE is in actuality not always the Haar ensemble. Interestingly, we find this is indeed the case for measurements along the $z$-basis ($\theta_m$\,$=$\,$0$):  
a projected state  $|\psi_A(\nu)\rangle$  is typically some {\it computational basis state} $|z_A\rangle$ with $z_A$\,$\in$\,$\{0,1\}^{N_A}$: \\
% \suggestion{--- clearly different from a Haar random vector. Theorem 2 below (with precise formulation in \cite{supp}) captures this}{}:\\
%
{\bf Theorem 2 (Informal).}  {\it
Let $|\psi_A(\nu)\rangle$\,$=$\,$\sum_{z_A} c_{z_A}(\nu) |z_A\rangle$ be the projected state corresponding to some fixed bit-string outcome $\nu$ 
 on RPS $\ket{\Psi_\pi}$ in the tilted-basis model with $\theta_0$\,$\notin$\,$\{0,\pi/2,\pi\}$ and $\theta_m$\,$=$\,$0$ ($z$-basis measurement). 
Then with unit probability over RPSs, only one coefficient $c_{z_A}(\nu)$ dominates in the thermodynamic limit.}

See \cite{supp} for a precise formulation of the Theorem. 
The key idea is that the Born weight $|c_{z_A}(\nu)|^2$ depends on the Hamming weight $h$ of the preimage bit-string $\pi^{-1}(z_A,\nu)$: $|c_{z_A}(\nu)|^2$\,$\propto$\,$\cos^{2(N-h)}(\theta_0/2)\sin^{2h}(\theta_0/2).$
For RPSs, these Hamming weights are asymptotically distributed as independently and identically distributed binomial variables in limit $N\to\infty$. Fluctuations of independent binomials typically produce an exponentially large separation between the largest and the next-largest Born weight, so with unit probability a single coefficient dominates and the post-measurement state collapses to a single  bit-string state $|\psi_A(\nu)\rangle \to |z_A(\nu)\rangle$.
Strictly speaking, Theorem 2 tells us the {\it average} behavior of a projected state over RPSs, but we expect  the same behavior for fixed {\it typical} RPSs. Furthermore, since the statement of Theorem 2 is agnostic as to which basis state is attained, it strongly suggests a uniform distribution over the computational basis on $A$. This defines the {\it classical bit-string ensemble} 
\begin{align}
\mathcal{E}_{\mathrm{Cl}}:=\bigl\{p(z_A)=1/2^{N_A}, |z_A\rangle\bigr\},
\end{align}
a discrete ensemble with the {\it least} entropy among the unravelings of $\rho_A$\,$=$\,$\mathbb{I}/d_A$, representing a maximal violation of deep thermalization~\cite{mark2024maximum, liu2024deep, Touil2024branchingstatesas, chang2025deep}.

 Is this failure of deep thermalization a fine-tuned feature of computational basis measurements, or does it represent a robust phase?
To probe this, we turn to numerics~(see the SM~\cite{supp} for details). We choose initial Bloch angles $(\theta_0,\phi_0)$\,$=$\,$(\pi/4,\pi/4)$ and generate PEs from RPSs, fixing $N_A$\,$=$\,$2$, then compute the second-moment trace distances  $\Delta^{(2)}_\text{Cl}$ and $\Delta^{(2)}_\text{Haar}$ from the classical bit-string and Haar ensembles respectively, varying the measurement angle $\theta_m$, see Fig.~\ref{Fig:1}(b). 
Indeed, convergence to $\mathcal{E}_\text{Cl}$ and $\mathcal{E}_\text{Haar}$ (exponentially fast in system size) is seen  for $\theta_m$\,$=$\,$0,\pi$ respectively. However, strikingly, we also observe a  critical value $\theta_m^*$\,$\approx$\,$0.193\pi$  demarcated by crossings of both trace-distance curves over different system sizes (higher moments show a similar behavior with the same critical point~\cite{supp}).
Below~(above) this value, the PE appears to converge to  $\mathcal{E}_\text{Cl}$ ($\mathcal{E}_\text{Haar}$), strongly indicating that the  absence of deep thermalization survives as a {\it robust} phase away from the $\theta_m$\,$=$\,$0$ point, separated  from a deeply-thermalized phase by a sharp phase transition  at $\theta^*_m$.

{\it Coherence-induced phase transition.}---What is the mechanism behind the transition in the PE and the nature of the accompanying phases?  
A key observation is that the apparent limiting ensembles $\mathcal{E}_\text{Cl}$ and $\mathcal{E}_\text{Haar}$ are distinguished by the ``amount'' of superposition over the computational basis  harbored by their constituent wavefunctions. 
%
% \matteo{sentence got lost here? reinstated}
This is captured by the resource-theoretic concept of {\it coherence}~\cite{baumgratz2014quantifying, PhysRevLett.116.120404, RevModPhys.89.041003}. As a resource, coherence can be rigorously quantified by the relative entropy of coherence $C_{r}(\rho)$\,$:=$\,$S(\rho_\text{diag})$\,$-$\,$S(\rho)$ (with $S$ the von Neumann entropy and $\rho_\text{diag}$ the diagonal part of $\rho$, corresponding to full dephasing), which is a monotone under ``free'' incoherent operations (operations which map  diagonal states to themselves). 
Among pure states, computational basis states achieve the minimum coherence, $C_{r}(|z_A\rangle)$\,$=$\,$0$, while Haar random states are close to maximally-coherent, 
% $C_{l_1}(|\psi_{A,\text{Haar}}\rangle) \approx \pi (d_A-1)/4 $. 
{$\mathbb{E}_{\psi_A \sim \text{Haar}(\mathcal{H}_A)}[C_{r}(|\psi_{A}\rangle)]$\,$=$\,$\sum_{k=2}^{d_A} 1/k $~\cite{singh2016average}}\footnote{Asymptotically, this grows as $\ln d_A + \gamma - 1$ with corrections in $1/d_A$, where $\gamma$ is the Euler-Mascheroni constant.}.
Furthermore, RPD and $z$-basis measurements are incoherent operations while general measurements are not (they can increase coherence). This suggests that the transition in the PE should be viewed as one driven by coherence:  coherence is injected globally by the initial state and choice of measurement basis before being scrambled nonlocally by RPD, which crucially does not create more superpositions;  %and then
such coherence 
may then either proliferate %to
and be detectable locally in the 
% local 
subsystem $A$ or not. 
% vanish;
Further, the picture implies this physics should generically occur, i.e., be universal, as long as the dynamics is sufficiently scrambling and {\it coherence-preserving}: indeed, the maximal set of such free incoherent unitary operations is precisely RPD, with the mild extension to include diagonal phase gates~(see \cite{supp} for details).
%
% \suggestion{the role of the RPD is to scramble the initial quantum information nonlocally,}{A coherence-preserving dynamics is therefore a necessary physical constraint for this mechanism~\cite{supp}. The role of the RPD is to scramble the initial quantum information nonlocally while simultaneously preserving coherence,} making this process universal. 
It also further suggests that the appropriate order parameter to consider is the ensemble-averaged coherence~\footnote{We emphasize that this is not the same as the coherence of the average state   $C_{r}([ \mathbb{E}_{\psi_A\sim\mathcal{E}_\text{PE}} |\psi_A \rangle \langle \psi_A| ] )$, since coherence is not a linear function of the state.} 
\begin{align}
\overline{C_{r}} := \mathbb{E}_{|\psi_A  \rangle\sim \mathcal{E}_\text{PE}}[C_{r}(|\psi_A\rangle) ].
\end{align}

\begin{figure}[t]
\includegraphics[width=0.5\textwidth]{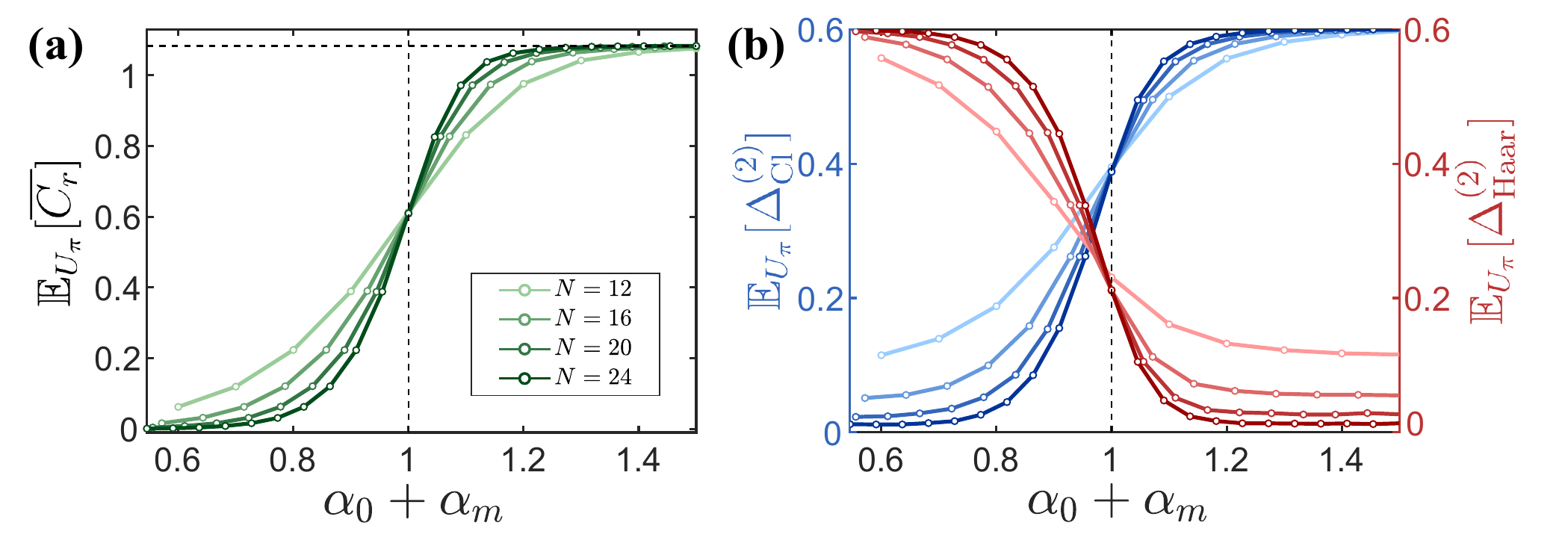}
\caption{(a) Ensemble-averaged coherence of the PE~($N_A$\,$=$\,$2$) for the mixed-basis model with $\alpha_0$\,$=$\,$0.5$, showing a  transition at $\alpha_m$\,$=$\,$0.5$~(vertical dashed line) as predicted. Horizontal dashed line indicates the coherence of Haar random states. (b) Trace distances of the PE~($k$\,$=$\,$2$)  confirming convergence to the classical bit-string and Haar ensembles in the deeply non-ergodic and ergodic regimes respectively. 
}
\label{Fig:2}
\end{figure}

To confirm this physical picture and illustrate it explicitly, we turn to the mixed-basis model. 
The key simplification of this model is that the distributions of computational-basis amplitudes of the initial state and measurement basis   are all ``flat'', allowing us to explicitly calculate coherence in the projected states through simple counting of the number of bit-strings appearing in its decomposition (this is unlike the general case where the shape of the distribution also matters).
As we will see, this yields an analytically predictable phase boundary. 

To begin, consider the RPS associated with the mixed-basis model, 
\begin{equation}
    \ket{\Psi_\pi} = 2^{-\alpha_0 N/2} \sum_{z\in S} (-1)^{f_1(z)}i^{f_2(z)} \ket{z}.
    \label{eqn:subset_state}
\end{equation}
Above, $S$ is the image of the bit-strings $0^{(1-\alpha_0)N} \times \{0,1\}^{\alpha_0 N}$ under $\pi$; this is a random subset of $\{0,1\}^N$ of cardinality $|S| = 2^{\alpha_0 N}$.  $f_1(z)$, $f_2(z)$ are Boolean functions determined by $\pi$, but may be treated as effectively independent pseudorandom functions. % when $\alpha_0<1$.
%\matteo{why $\alpha_0<1$?}.
% \ww{not sure, it is just my feeling that if $\alpha_0$ is too big, the independent pseudorandomness is weaker and weaker, but let's drop since not precise}
We note Eq.~\eqref{eqn:subset_state}  is a generalized version of so-called {\it random subset-phase states}~\footnote{Generalized to include imaginary phases}, which have been recently studied  in the context of pseudoentanglement~\cite{aaronson2022quantum,giurgica2023pseudorandomness,jeronimo2023pseudorandom,Feng2025}. 

Next, we analyze the form of a given projected state. Assume we obtain the measurement outcome $\nu$\,$=$\,$00\cdots 0$.
The associated local post-measurement state  $|\psi_A(\nu)\rangle$ is  proportional to
\begin{align}
 \sum_{z\in S} (-1)^{f_1(z)}i^{f_2(z)}\left( \prod_{j=N_A+1}^{N_A+(1-\alpha_m) N_B} \delta_{z_j,0} \right) \ket{z_A(z)},
 \label{eqn:toy_superposition}
 \end{align}
where $z_A(z)$\,$=$\,$(z_1$\,$,$\,$\cdots$\,$,$\,$z_{N_A})$ is the restriction of the global bit-string $z$ to the subsystem $A$. 
 The Kroenecker deltas in Eq.~\eqref{eqn:toy_superposition} come from imposing compatibility with the $z$-basis measurement outcomes on $(1-\alpha_m)N_B$ qubits; for the remaining $\alpha_m N_B$ qubits measured in the $x$-basis, there is no such constraint. 
For other measurement outcomes $\nu$, the projected states have  similar forms upon suitably redefining the ``pseudorandom'' Boolean functions $f_{1,2}(z)$ and delta functions. 

We now ask about the number $l_A$ of nonzero terms which contribute to the sum in Eq.~\eqref{eqn:toy_superposition}. 
This can be estimated as follows: the initial state provides $|S|$\,$=$\,$2^{\alpha_0 N}$ bit-strings; this number gets approximately halved by each Kronecker delta for a total of $(1-\alpha_m)N_B$ times, resulting in $l_A$\,$\sim$\,$2^{\alpha_0 N_A + (\alpha_0 + \alpha_m - 1)N_B}$. 
Therefore,  $\alpha_0$\,$+$\,$\alpha_m$\,$=$\,$1$ demarcates the boundary between two distinct behaviors as $N_B$\,$\to$\,$\infty$: for $\alpha_0$\,$+$\,$\alpha_m$\,$<$\,$1$, the expected number of terms in the sum vanishes. Of course, the Born rule ensures only nonvanishing states are selected,  so in practice $l_A$\,$\sim$\,$1$. As long as $\alpha_0 N$\,$>$\,$N_A$, the PE covers all bit-strings $z_A$ uniformly, thus giving  the classical bit-string ensemble $\mathcal{E}_\text{Cl}$. 
Conversely, for $\alpha_0$\,$+$\,$\alpha_m$\,$>$\,$1$, $l_A$ diverges. Each bit-string $z_A$ acquires many contributions with independent  random coefficients $\pm$\,$1$\,$,$\,$\pm$\,$i$. By the central limit theorem, this sum (upon  normalization) converges in distribution to a complex standard Gaussian random variable, which describes a Haar random state. %  on $A$. 
This collection of states is thus the Haar ensemble $\mathcal{E}_\text{Haar}$. (We note that a {\it complex} Haar random vector arose because of the initial state $|\mathrm{Y}_+\rangle$ which provides imaginarity~\cite{hickey2018quantifying,wu2021resource}; if instead it was real like $|\mathrm{X}_\pm\rangle$\,$=$\,$\frac{1}{\sqrt{2}}(|0\rangle\pm|1\rangle)$ we would obtain a {\it real} Haar random vector, i.e., $\mathcal{E}_\text{Haar}$ would be  the state ensemble invariant under the real orthogonal group. See \cite{supp} for more details.)

To verify these predictions, we numerically evaluate the ensemble-averaged  relative entropy of coherence $\overline{C_{r}}$ for the mixed-basis model with $\alpha_0$\,$=$\,$0.5$~(see \cite{supp} for simulations with other values of $\alpha_0$). 
Fig.~\ref{Fig:2}(a) demonstrates an unambiguous transition at  $\alpha_m$\,$=$\,$0.5$ which becomes sharper  with increasing $N_B$; 
finite-size scaling shown in the SM~\cite{supp} is consistent with a first-order transition, which fits well with the analytical derivation presented above. 
% 
% finite-size scaling shown in the SM~\cite{supp} support the first-order nature of the transition suggested by the microscopic analysis.  
%
Further, Fig.~\ref{Fig:2}(b) shows convergences to the expected limiting ensembles $\mathcal{E}_\text{Cl}$ and $\mathcal{E}_\text{Haar}$ flanking  the transition. This thereby provides strong evidence that both the tilted and mixed-basis models exhibit the same deep ergodicity-breaking physics, 
with coherence as the underlying driver behind the singular change in their PEs. 
%\suggestion{Indeed,  this connection is further strengthened by an estimation of the critical point of the tilted-basis model from the mixed-basis model, shown in the End Matter, yielding a prediction $\theta^{*'}_m \approx 0.181\pi$, consistent with numerics.}
This connection can be probed quantitatively: in the End Matter we explain how one may convert mixed-basis parameters $(\alpha_0,\alpha_m)$ to tilted-basis parameters $(\theta_0, \theta_m)$ of equal coherence (as measured by $C_r$); thus $\alpha_0$\,$+$\,$\alpha_m$\,$=$\,$1$ gives a prediction for the tilted-basis phase boundary. For the data in Fig.~\ref{Fig:1}(b) this yields $\theta^{*'}_m$\,$\approx$\,$0.181\pi$, reasonably close to the observed $\theta^{*}_m$\,$\approx$\,$0.193\pi$. However, we do not expect this connection to be exact, and indeed critical properties in the two models may be different---seemingly first-order (second-order) for the mixed (tilted) basis model; see~\cite{supp}. Further study of the critical properties is an interesting direction for future work.

% However, we do not expect this connection to be  exact: though we show the transitions in both models are continuous \CL{first order transition} through  finite-size scaling of the order parameter (see  \textcolor{red}{Appendix F}), they  are characterized by different critical exponents. 

% Together with the tilted-basis model, our analyses and numerical simulations demonstrate that the deep ergodicity-breaking phase of the PE, characterized by 
% Fig.~\ref{Fig:2} also shows the ensemble-averaged coherence computed for the RPS model shown in Fig.~\ref{Fig:1}(b) which yields a consistent critical point, demonstrating the faithfulness and effectiveness of the toy model in capturing the essential physics behind the transition. 
% In the \textcolor{red}{End Matter}, we provide further evidence of this by relating in a loose but quantitative fashion the critical point of the RPS model to  the toy model through the inverse participation ratio (IPR) of the initial states~(a proxy for the $l_1-$coherence); further, 

Lastly, to bolster %the 
our argument of 
universality of coherence-induced deep thermalization  phases beyond the microscopic models studied here, we perform in the   SM~\cite{supp} an independent analytical calculation of the ensemble-averaged inverse participation ratio~(IPR), a proxy for coherence, beginning from {\it generic} RPSs and under measurement schemes with varying IPRs.
While based on an approximation  replacing ``quench averaging'' over measurement outcomes   with ``annealed averaging'', it reveals distinct scalings in the IPR consistent again with the existence of deeply ergodic and non-ergodic phases. 
% All together, these indicate the coherence-induced phase transition in the PE we have uncovered is a universal, generic phenomenon of quantum dynamics within the RPD class. 

{\it Discussion and outlook.}---Our work has identified a novel form of ergodicity-breaking transition in quantum many-body dynamics, defined not by a failure of conventional thermalization, but of deep thermalization.
Importantly, this phenomenon is invisible in expectation values of standard observables,
unlike known examples of ergodicity breaking such as localization~\cite{basko2006metal,gornyi2005interacting,nandkishore2015many,alet2018many,abanin2019colloquium,oganesyan2007localization,vznidarivc2008many} or many-body scarring~\cite{bernien2017probing,moudgalya2018entanglement,turner2018weak,choi2019emergent,ho2019periodic,lin2019exact,scherg2021observing,su2023observation,bhowmick2025asymmetric}. Rather, it is similar to the phenomenon of measurement-induced phase transitions~\cite{li2018quantum,li2019measurement,skinner2019measurement,chan2019unitary,bao2020theory, annurev:/content/journals/10.1146/annurev-conmatphys-031720-030658}, where individual quantum trajectories exhibit transitions not detectable within the average state.
From a quantum information theoretic perspective, our transition can also be  understood as a sharp change in the informational content of the PE: the accessible information of the limiting Haar (classical bit-string) ensemble is minimal (maximal), saturating  the subentropy (Holevo) bounds~\cite{jozsa1994lower,Nielsen_Chuang_2010,holevo1973bounds}. 

We have further  identified the transition as being driven by coherence, a quantum resource which is preserved by the dynamics, but is injected through the choice of initial states and measurement basis and transferred to the local unmeasured subsystem. 
This framework readily lends itself to generalizations based on other information-theoretic resources, such as imaginarity~\cite{hickey2018quantifying,wu2021resource}, non-stabilizerness~\cite{PhysRevA.86.052329, PhysRevLett.118.090501, PRXQuantum.3.020333, LoioMagic2025} or non-Gaussianity~\cite{PhysRevA.97.052317, PhysRevA.98.022335}.  Do resources injected by input states and measurements, and scrambled by ``free'' dynamics, generally give rise to PE transitions between a ``resourceful'' deeply-thermalized phase and a ``resourceless'' ergodicity-breaking phase? How do such transitions, which require the constraint of dynamics being resource-preserving, interplay with conventional symmetry constraints like energy or charge conservation?
%It would also be highly interesting to understand how such resource-induced deep thermalization transitions interplay with standard conservation laws in Hamiltonian dynamics. %On the other hand, how do such resource-induced deep thermalization transitions interplay with standard conservation laws in generic Hamiltonian dynamics? 
Charting the landscape of resource-induced deep thermalization transitions, their implications for many-body dynamics, and possible applications in quantum information science represent exciting directions for future research.
% Identifying such resources and constructing explicit models exhibiting such transitions to explore the landscape of distinct universality classes within the `deep' Hilbert space will no doubt \matteo{always some doubt haha} be exciting lines of directions for future work. 
% \matteo{how about: charting the landscape of resource-induced deep thermalization transitions, their implications for many-body dynamics, and possible applications in quantum information science represent exciting directions for future research.} 

%Moving forward, uncovering more generally the landscape of the distinct universality classes that the PE can exhibit and the nature of the phase transitions between them, will no doubt be exciting lines of directions for future work. 
%\ww{this discussion about imaginarity is not quite right...  this is more restricting dynamics.Breaking imaginarity always leads to complex phases, so realness does not survive as a phase.need to rewrite.}

\begin{acknowledgments}
{\it Acknowledgments}.---We thank Andreas W.~W.~Ludwig and Tianci Zhou for helpful discussions, and Daniel Mark for useful comments. 
M.~I.~is supported by the U.S. Department of Energy, Office of Science, Office of Advanced Scientific Computing Research under Award Number DE-SC0025615. 
W.~W.~H.~is supported by the Singapore National Research Foundation (NRF) Felllowship NRF-NRFF15-2023-0008 and through the National Quantum Office, hosted in A*STAR, under its Centre for Quantum Technologies Funding Initiative (S24Q2d0009). 
This research was supported in part by grant NSF PHY-2309135 to the Kavli Institute for Theoretical Physics (KITP). 
\end{acknowledgments}

\bibliography{biblio}

 \section{End matter}

 {\it Local random permutation dynamics (RPD) model.}---In the main text we analyzed initial states $|\Psi_0\rangle$ evolved under the application of a single {\it global} random permutation unitary $U_\pi$, where $\pi \in S_{2^N}$ (``global RPDs''). 
 %We claimed that this models the dynamical behavior  of deep circuits made of random {\it local} permutation gates (``local RPDs''). 
Here, we numerically confirm that the deep thermalization phase transition identified in the main text is indeed unaffected upon considering ``local RPDs'' --- deep circuits made of random {\it local} permutation gates.

Concretely, we consider quantum circuits composed of local permutation unitaries, each of which acts on a fixed number of sites $r$ and  stacked in a brickwork pattern up to some depth $t$, see e.g.,  Fig.~\ref{Fig:1}(a) for $r=3$.
Technically, such ``$r$-local'' RPDs cannot reproduce exactly the distribution of global permutation unitaries for any  $r<N$ even as $t \to \infty$, since  they cannot generate {\it odd} permutations $\pi \in S_{2^N}$. 
However, for all $r\geq 3$, $r$-local RPDs generate the whole subgroup of even permutations (the {\it alternating group})~\cite{shende2003synthesisreversible}, which forms a $k$-design over the global permutation group for all $k \leq 2^N-2$, see~\cite{gowers1996almost,chen2024incompressibility,gay2025pseudorandomness}. (The cases $r=1,2$ are special as they only give Pauli and Clifford gates respectively.) As the discrepancy between deep $(r\geq 3)$-local RPDs and global RPDs arises only at exponentially high moments, we expect it to be unimportant toward the universal deep thermalization behavior.

\begin{figure}[t]
\includegraphics[width=0.5\textwidth]{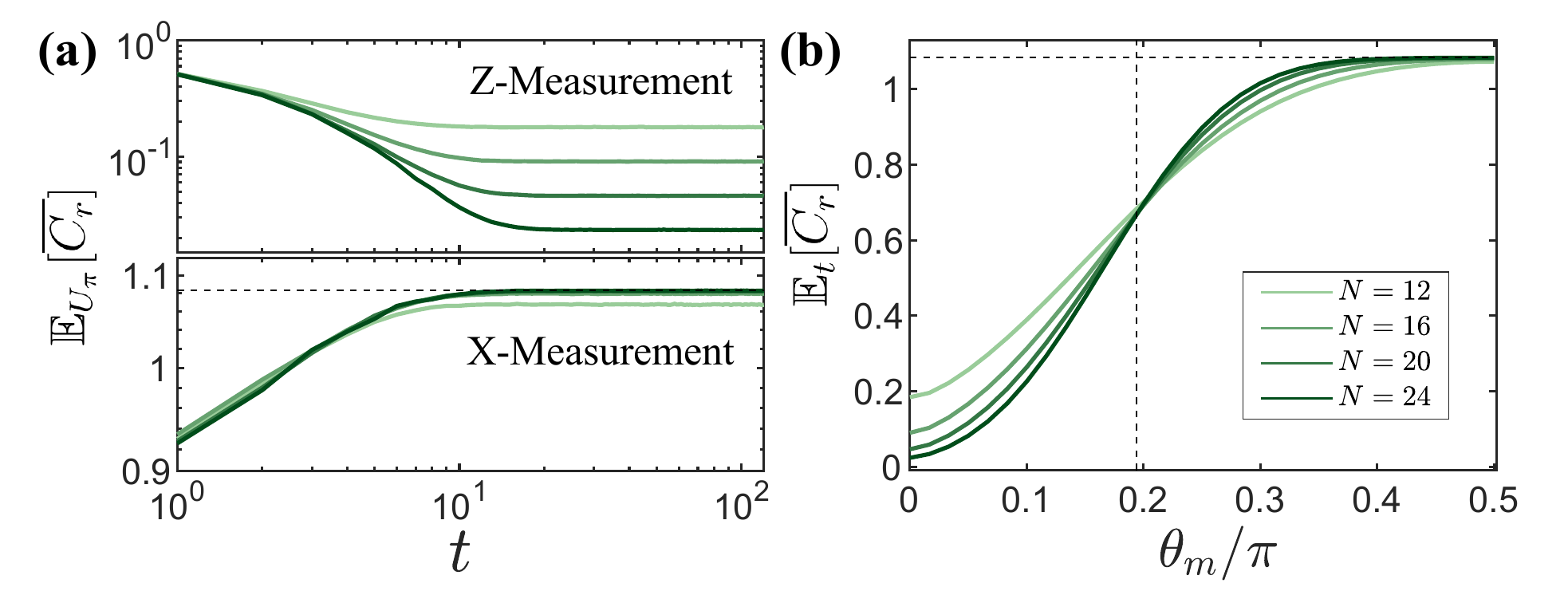}
\caption{ (a) Time evolution under 3-local RPDs of the ensemble-averaged coherence in the tilted-basis model, for $z$- ($x$-) basis measurements, shown in the top (bottom) panel. In both cases, the coherence saturates to that of the classical bit-string ensemble and Haar ensemble respectively. %The global state is produced by a depth-$t$ circuit of local 3-qubit random permutation gates, starting from a state with Bloch angle $(\theta$\,$,$\,$\phi)$\,$=$\,$(\pi/4$\,$,$\,$\pi/4)$.
(b) Saturation value  as a function of the measurement angle $\theta_m$. %The saturation time scales approximately as $\sim N$, and the time-average is 
% For each system size $N$, we take the late-time saturation values to be from times $t\in[2N,4N]$. 
The critical angle $\theta_m^*$\,$\approx$\,$0.193\pi$ of the tilted-basis model from Fig.~\ref{Fig:1}(b) is shown as a vertical dashed line. 
% The horizontal dashed line indicates the coherence of Haar random states.
}
\label{Fig:3}
\end{figure}

To verify this, we numerically simulate $3$-local RPDs within the tilted-basis model, choosing initial Bloch angles $(\theta_0$\,$,$\,$\phi_0)$\,$=$\,$(\pi/4$\,$,$\,$\pi/4)$. We construct the PE for a subsystem of size $N_A$\,$=$\,$2$ and track its ensemble-averaged coherence during circuit time evolution. 
Fig.~\ref{Fig:3}(a) shows that, for both $z$-basis and $x$-basis measurements, the ensemble-averaged coherence converges in time to that of the corresponding limiting ensembles, with errors vanishing in the thermodynamic limit.
In Fig.~\ref{Fig:3}(b) we plot the saturation value of the coherence, taken to be values in times $t \in [2N,4N]$, versus $\theta_m$, which reproduces very well the same deep thermalization phase transition behavior seen for global RPD in Fig.~\ref{Fig:1}(b).

{\it Estimation of critical point of the tilted-basis model from the mixed-basis model.}---Here, we estimate the critical point of the tilted-basis model  by using our analytical result for the mixed-basis model, $\alpha_0+\alpha_m=1$, and matching the amount of coherence in the initial states and in the measurement bases of the two models.

We convert between model parameters by using the relative entropy of coherence $C_r$. For any pure state $|\Psi\rangle$, the latter reduces to the Shannon entropy $S(\{p_z\}):=-\sum_z p_z\ln p_z$ of its population over  the computational basis,
\begin{equation}
    C_r(\Psi) = S(\{p_z\}),
\end{equation}
with $p_z = |\langle z|\Psi\rangle|^2$. 
 Below we denote the initial state by $|\Psi_0\rangle$ and the measurement-basis state by $|\Psi_m\rangle$.

In the tilted-basis model the bit-string distribution is a sum of independent binomials, so  
\begin{align}
    &C_r(\Psi_{0})=NH_2(\cos^2(\theta_0/2)), \nonumber \\
    &C_r(\Psi_m) = N_B H_2(\cos^2(\theta_m/2)),
\end{align}
 with $H_2$ the binary Shannon entropy $-p\ln(p)-(1-p)\ln(1-p)$, while for the mixed-basis model
 \begin{equation}
     C_r(\Psi_0) = \alpha_0 N \ln2,\quad C_r(\Psi_m) = \alpha_m  N_B \ln2.
 \end{equation}
 Equating the coherences of the input and measurement states between the two models yields $\alpha_0,\alpha_m$ given $\theta_0,\theta_m$;  
 % , solving for effective $\alpha_0(\theta_0), \alpha_m(\theta_m)$ for the tilted-basis model. 
  applying then the threshold condition $\alpha_0+\alpha_m=1$ gives
 \begin{equation}
     H_2(\cos^2(\theta_0/2)) + H_2(\cos^2(\theta_m/2)) = \ln 2.
 \end{equation}

For the parameters used in the main text, $(\theta_0$\,$,$\,$\phi_0)$\,$=$\,$(\pi/4$\,$,$\,$\pi/4)$, solving for $\theta_m$ yields a prediction $\theta^{*'}_m \approx 0.181\pi$, which is close to the numerically extracted transition  of $\theta_m^*\approx 0.193\pi$. 
A discrepancy is to be expected as the mixed-basis model can be thought of as a simplification of the tilted-basis amplitude distribution (effectively truncating amplitudes to $0$ or $1$). 
Nevertheless, the fact that these two values match well quantitatively is indication of the common physics underlying the two models.

\end{document}

\title{Supplemental material for: 
\\
Coherence-induced deep thermalization transition in  random permutation quantum dynamics}
\author{Chang Liu}
\affiliation{Department of Physics, National University of Singapore, Singapore 117551}
\author{Matteo Ippoliti}
\affiliation{Department of Physics, The University of Texas at Austin, Austin, TX 78712, USA}
\author{Wen Wei Ho}
\affiliation{Department of Physics, National University of Singapore, Singapore 117551}
\affiliation{Centre for Quantum Technologies, National University of Singapore,   Singapore 117543}

%@misc{supp,note ={ See Supplemental material online for details of statements and proofs presented in the main text, whichincludes Refs.~\cite{}. } }

\date{\today}
\maketitle

In this supplemental material, we provide  proofs supporting the theorems stated in the main text,  
as well as details on analyses of the models and  numerical simulations. It is organized as follows. In Appendix~\ref{sec:preliminaries}, we review Weingarten calculus on the symmetric group, the mathematical machinery underlying almost all of our rigorous calculations. In Appendix~\ref{sec:regular_thermalization}, we prove the statement (Theorem~1 from the main text) that regular thermalization occurs generically for random permutation states in the thermodynamic limit.  In Appendix~\ref{sec:eq_class_theory}, we apply the maximum entropy principle of \cite{chang2025deep} described in the main text to the tilted-basis model, to derive that the generalized Scrooge ensemble (which is the predicted limiting form of the projected ensemble predicted by the principle) is always the Haar ensemble, regardless of measurement direction. 
In Appendix~\ref{sec:PE_RPS_analytic}, we compute the limiting form of the projected ensemble of the tilted-basis model for the specific cases of $z$- and $x$-basis measurements, and in particular prove Theorem~2 of the main text. We also provide additional details on numerical simulations performed for  the tilted basis model including other measurement directions, determination of the critical point, and finite-size scaling collapse in Appendix~\ref{sec:PE_RPS_numerics}. %%
In Appendix~\ref{sec:toy_model}, we present further details of the mixed-basis models, including rigorous statements on regular thermalization akin to Theorem 1 in the tilted-basis model, as well as additional numerical results like finite-size collapse.
In Appendix~\ref{sec:imaginarity}, we consider the case when input states, dynamics, and measurement bases are all real: this results in the emergence of the real orthogonal Haar ensemble in the deep thermalized phase. 
% 
%Finally in Appendix~\ref{sec:IPR}, we present an argument for the universality of the coherence-induced deep thermalization phase transition for arbitrary input states and measurement bases, by  computing and examining the scaling behavior of the ensemble-averaged inverse participation ratio, which reveals the existence of two distinct phases.
In Appendix~\ref{sec:IPR}, we present an argument for the universality of the coherence-induced deep thermalization phase transition for arbitrary input states and measurement bases, by computing and examining the scaling behavior of the ensemble-averaged inverse participation ratio, which reveals the existence of two distinct phases. Finally, in Appendix~\ref{sec:monomial}, we discuss the most general class of coherence-preserving unitaries, providing additional numerical results for circuit models composed of local random permutation gates with diagonal phase gates and showing that the same coherence-induced deep thermalization phenomenon occurs.
% of the post-measurement state and demonstrate the existence of two distinct phases. 
% In Appendix~\ref{sec:imaginarity}, we analyze another resource, imaginarity, which also captures the phase transition; moreover, we use imaginarity to discuss the realness of the projected ensemble. 
% Finally, in Appendix~\ref{sec:phase transition}, we examine the nature of the deep thermalization phase transition, including an error analysis of the critical point and evidence suggesting that it may be a first-order phase transition. \ww{let's weaken the claim of nature of the phase transition, leave it open-ended}

\appendix
\section{Mathematical preliminaries}
\label{sec:preliminaries}
\subsection{Weingarten calculus on the symmetric group}

In the main text we study random permutation dynamics, focusing  focusing on two microscopic models that both involve the application of a global random permutation unitary. To analyze the behavior of subsystems and projected ensembles under such dynamics, it is useful to perform averages over such unitaries; this entails Weingarten calculus on the symmetric group. Here we establish  basic notations used and refresh the reader on the mathematics involved.

Consider an $N$-qubit system with Hilbert space dimension $d = 2^N$. A permutation unitary $U_{\pi}$ is a $d\times d$ matrix with exactly one nonzero entry (equal to 1) in each row and column, acting classically on computational basis states as
\begin{equation}
    U_\pi|z\rangle = |\pi(z)\rangle,
\end{equation}
where $\pi$ is a permutation from the symmetric group $S_d$, and $|z\rangle$ is a  basis state with $z\in\{0,1\}^N$.
A random permutation unitary is one which is drawn uniformly from the $d!$ elements of $|S_d|$. 

Next consider the Haar average of the $2m$-fold tensor product over the symmetric group $S_{d}$. Since $S_{d}$ is finite and each $U_{\pi}$ is  real, we have
\begin{equation}
     \mathbb{E}_{U_\pi} [U_{\pi}^{\otimes m}\otimes U_{\pi}^{* \otimes m}] = \frac{1}{d!}\sum_{\pi\in S_{d}} [U_{\pi}^{\otimes m}\otimes U_{\pi}^{\otimes m}] = \sum_{i,j}\mathrm{Wg}^{(2m)}(\sigma_i,\sigma_j;d)|\sigma_i\rangle\langle\sigma_j|,\label{Haar_permutation}
\end{equation}
where $\mathrm{Wg}$ denotes the Weingarten function, and $\sigma_i$ labels partitions of the index set $Z_{2m}=\{1,2,\ldots,2m\}$. Here, a partition $\sigma$ of $Z_{2m}$ is a set of pairwise disjoint nonempty subsets of $Z_{2m}$ whose union covers the entire set. For instance, the set $Z_2=\{1,2\}$ admits two partitions: $\sigma_1=\{\{1,2\}\}$ and $\sigma_2=\{\{1\},\{2\}\}$ .

The operator in Eq.~\ref{Haar_permutation} acts as a projector onto the subspace spanned by the permutation-invariant basis $\{|\sigma_i\rangle\}$. Note that the set of $m$-th Haar-invariant basis elements $\{\tau_j\}$ forms a subset of $\{|\sigma_i\rangle\}$, where each $\tau_j$ corresponds to a permutation acting on $\mathcal{H}^{\otimes m}$.  

Here, we list all basis states $\{|\sigma_i\rangle\}$ for $2m=2$ and $4$, which will be used in the following proofs.

For $2m=2$, we have:
\begin{equation}
    |\sigma_1\rangle = \sum_{a=1}^{d} |a,a\rangle, \quad |\sigma_2\rangle = \sum_{a,b=1}^{d} |a,b\rangle.\label{permutation_basis1}
\end{equation}

For $2m=4$, there are 15 partitions, giving the corresponding basis states:

\begin{align}
    &|\sigma_1\rangle = \sum_{a=1}^{d} |a,a,a,a\rangle, \quad |\sigma_2\rangle = \sum_{a,b=1}^{d} |a,a,a,b\rangle,
\quad |\sigma_3\rangle = \sum_{a,b=1}^{d} |a,a,b,a\rangle,
\quad |\sigma_4\rangle = \sum_{a,b=1}^{d} |a,b,a,a\rangle,\nonumber\\
& |\sigma_5\rangle = \sum_{a,b=1}^{d} |a,b,b,b\rangle,
\quad |\sigma_6\rangle = \sum_{a,b=1}^{d} |a,a,b,b\rangle,
\quad |\sigma_7\rangle = \sum_{a,b=1}^{d} |a,b,b,a\rangle,
\quad |\sigma_8\rangle = \sum_{a,b=1}^{d} |a,b,a,b\rangle,\nonumber\\
&|\sigma_9\rangle = \sum_{a,b,c=1}^{d} |a,a,b,c\rangle,
\quad |\sigma_{10}\rangle = \sum_{a,b,c=1}^{d} |a,b,a,c\rangle,
\quad |\sigma_{11}\rangle = \sum_{a,b,c=1}^{d} |a,b,c,a\rangle,
\quad |\sigma_{12}\rangle = \sum_{a,b,c=1}^{d} |a,b,c,b\rangle,\nonumber\\
& |\sigma_{13}\rangle = \sum_{a,b,c=1}^{d} |a,b,b,c\rangle,
\quad |\sigma_{14}\rangle = \sum_{a,b,c=1}^{d} |a,b,c,c\rangle,
\quad |\sigma_{15}\rangle = \sum_{a,b,c,d=1}^{d} |a,b,c,d\rangle.\label{permutation_basis2}
\end{align}

%In our proofs, 
 %%%%%%%%%%%%%%%%%%%%%%%%%
In the following, we will also use the asymptotic form of the Weingarten function for the symmetric group, as given in Corollary 2.3 of~\cite{collins2025weingarten}:
\begin{equation}
    \mathrm{Wg}^{(k)}(\sigma_1,\sigma_2; d) 
    = \mu(\sigma_1\wedge \sigma_2, \sigma_1)\, \mu(\sigma_1 \wedge \sigma_2, \sigma_2)\, d^{-\#(\sigma_1 \wedge \sigma_2) } \left(1 + O(d^{-1})\right),\label{asymptotic_Wg}
\end{equation}
where $\#\sigma$ denotes the number of blocks in the partition $\sigma$, and $\mu$ is the M{\"o}bius function on the lattice of partitions.  The operation $\wedge$ denotes the \emph{meet} of two partitions: for a finite set $Z$ with partition space $\mathcal{P}(Z)$, and $\sigma_1,\sigma_2 \in \mathcal{P}(Z)$, the meet $\sigma_1 \wedge \sigma_2$ is defined as the finest partition that is coarser than both $\sigma_1$ and $\sigma_2$, i.e., their greatest lower bound in the lattice of partitions.

\section{Regular thermalization in the tilted-basis model}
\label{sec:regular_thermalization}

In this section, we provide the proof of Theorem 1 in the main text, which establishes that regular thermalization occurs for a typical (i.e., excluding some measure zero set) random permutation state in the tilted-basis model. We restate Theorem 1 here:

{\bf Theorem 1.} {\it 
Let $|\Psi_\pi\rangle$ be an $N$-qubit RPS in the tilted-basis model specified by an initial state $|\Psi_0\rangle$ with Bloch-sphere angles $(\theta_0,\phi_0)$ not in $\{\theta_0$\,$=$\,$0,\theta_0$\,$=$\,$\pi,(\theta_0$\,$=$\,$\tfrac{\pi}{2},\phi_0$\,$=$\,$0)\}$. For any $\epsilon$\,$>$\,$0$, 
\begin{align}
\mathbb{P}\left(\left\|\rho_A - \frac{\mathbb{I}_A}{d_A}\right\|_1 > \epsilon\right) 
< \frac{(d^{-\alpha_0}+ d^{-2\beta_0})C+ d^{-1}D}{\epsilon^2},
%< \frac{1}{\epsilon^2} \left[ (d^{-\alpha_0}+ d^{-2\beta_0})C+ d^{-1}D \right],
%\left[ d^{-\alpha_0}(d_A - 1) + d^{-2\beta_0}(d_A - 1) + d^{-1}(d_A^2 - d_A + 1) \right],
%\label{theorem1}
\end{align}
where % $d_A = 2^{N_A} \ll d = 2^N$
$d_A=2^{N_A},d=2^N$ are the Hilbert-space dimensions of $A$ and the full system respectively, 
$0 < \alpha_0, \beta_0 < 1$ are factors determined by $(\theta_0,\phi_0)$, 
and $C,D$ are $O(1)$ in system-size factors. 
}

\textit{Proof.} We begin by relating the deviation of $\rho_A$ from the maximally mixed state to the ensemble-averaged purity.  
For any $d_A \times d_A$ matrix $M$, one has $\|M\|_1 \leq \sqrt{d_A}\,\| M \|_2$, where $\|\cdot\|_2$ denotes the Frobenius norm. This gives
\begin{align}
\mathbb{P}(\|\rho_A -\mathbb{I}_A/d_A\|_1 > \epsilon) \leq \mathbb{P}(\|\rho_A -\mathbb{I}_A/d_A\|_2 > \epsilon/\sqrt{d_A}) = \mathbb{P}(\|\rho_A -\mathbb{I}_A/d_A\|_2^2 > \epsilon^2/{d_A}).
\end{align}
Noting that $\|\rho_A -\mathbb{I}_A/d_A\|_2^2$ is a non-negative random variable, we invoke Markov's inequality to further bound
\begin{align}
\mathbb{P}(\|\rho_A  -\mathbb{I}_A/d_A\|_1 > \epsilon) & <  \frac{d_A   }{\epsilon^2}\mathbb{E} \|\rho_A  -\mathbb{I}_A/d_A\|_2^2 \nonumber \\
& = \frac{d_A   }{\epsilon^2}\text{Tr}\left(\mathbb{E} [\rho_A ^2] - 2 \mathbb{E} [\rho_A ]/d_A + \mathbb{I}_A/d_A^2 \right) \nonumber \\
& = \frac{d_A   }{\epsilon^2} \left( \text{Tr}(\mathbb{E} [\rho_A ^2]) - 2 \text{Tr} (\mathbb{E} [\rho_A ])/d_A + 1/d_A \right) \nonumber \\
& = \frac{d_A   }{\epsilon^2} \left( \text{Tr}\mathbb{E} ([\rho_A ^2]) -  1/d_A \right), \label{thermalization_bound}
\end{align}
where $\mathbb{E}[\cdot]$ denotes  ensemble-averaging.
This bound is general and does not depend on the choice of initial state $|\Psi_0\rangle$ or the averaging ensemble.% $\mathcal{U}$.  

We now specialize to the random permutation states of the tilted-basis model
\begin{equation}
    |\Psi_{\pi}\rangle = U_{\pi} |\Psi_0\rangle,\label{RPS}
\end{equation}
where the initial product state is
\begin{equation}
    |\Psi_0\rangle = \left(\cos(\theta_0/2)|0\rangle + e^{i\phi_0}\sin(\theta_0/2))|1\rangle\right)^{\otimes N}.
\end{equation}
To evaluate the purity, we require the four-fold tensor product of $U_\pi$. Using  Weingarten calculus and the permutation-invariant basis states in Eq.~\eqref{permutation_basis2}, one finds~\cite{szasz2025entanglement} %cite page curve paper
\begin{align}
    \mathbb{E}_{U_\pi}[\text{Tr}[(\rho_A^2)] &=\text{Tr}(S_A\mathbb{E}_{U_\pi}[\rho_A \otimes \rho_A])\\ \nonumber &= \text{Tr}(S_A\mathbb{E}_{U_\pi}([U_\pi|\Psi_0\rangle\langle\Psi_0| U_\pi^\dagger)^{\otimes 2}_A])\\ \nonumber
    &= \sum_{i,j} \mathrm{Wg}^{(4)}(\sigma_i,\sigma_j)\langle S_A\otimes I_B|\sigma_i\rangle \langle\sigma_j|(\Psi_0\otimes\Psi_0^*)^{\otimes 2}\rangle\\ \nonumber
    &=1/d_A+g^{N}(1-1/d_A)+1/d(1-g^N)(d_A-2+1/d_A)+(1-1/d_A)(f^{2N}/d^2)
    +o(1/d)\\ \nonumber
    &<1/d_A+g^{N}(1-1/d_A)+1/d(d_A-1+1/d_A)+(1-1/d_A)(f^{2N}/d^2),
\end{align}
where $S_A$ and $I_B$ denote the swap operator on subsystem $A$ and the identity on subsystem $B$ in the doubled Hilbert space $\mathcal{H}^{\otimes 2}$, respectively.  
Here,
\begin{align}
    g^N &= \langle \sigma_1|(\Psi_0\otimes\Psi_0^*)^{\otimes 2}\rangle = (\sin^4{(\theta_0/2)}+\cos^4{(\theta_0/2)})^N, \\
   f^{2N} &= \langle \sigma_{15}|(\Psi_0\otimes\Psi_0^*)^{\otimes 2}\rangle\rangle = (1+\sin{\theta_0}\cos{\phi_0})^{2N},
\end{align}
with $g^N$ representing the inverse participation ratio (IPR) of the initial state. 

Since $0 < g \leq 1$ and $0 \leq f \leq 2$, we may write $g = 2^{-\alpha_0}$ and $f = 2^{1 - \beta_0}$ with $0 \leq \alpha_0,\beta_0 \leq 1$.  
Substituting these scalings into the probability bound gives
\begin{align}
\mathbb{P}(\|\rho_A -\mathbb{I}_A/d_A\|_1 > \epsilon) & <  \frac{d_A   }{\epsilon^2} \left( \text{Tr}\mathbb{E}_{U_\pi} [\rho_A(U)^2] -  1/d_A \right)\\
&<(2^{-\alpha_0N}(d_A-1)+1/d(d_A^2-d_A+1)+2^{-2\beta_0N}(d_A-1))/\epsilon^2 \\
&= (d^{-\alpha_0}(d_A-1)+d^{-2\beta_0}(d_A-1)+ d^{-1} (d_A^2-d_A+1))/\epsilon^2\\
&=((d^{-\alpha_0}+ d^{-2\beta_0})C+ d^{-1}D)/\epsilon^2,\label{regular_thermalization}
\end{align}
where constants $C = (d_A-1)$, and $D = (d_A^2-d_A+1)$.

However, the bound above is trivial in the following exceptional cases:

\begin{enumerate}
  \item $\theta_0 = 0$ or $\pi$, i.e., $\alpha_0 = 0$: the initial state is a computational-basis product state; a  permutation unitary merely permutes basis strings and no thermalization occurs.
  \item $\theta_0 = \pi/2$ and $\phi_0 = 0$, i.e., $\beta_0 = 0$: the initial state is $|+\rangle^{\otimes N}$, which is invariant under permutation. Consequently, the state remains unchanged and no thermalization takes place.
\end{enumerate}

Excluding these parameter values, the bound in Eq.~\eqref{regular_thermalization} vanishes in the thermodynamic limit for any fixed subsystem dimension $d_A$. Therefore, we establish a sufficient condition for regular thermalization in random permutation states.  This concludes the proof of Theorem~1. $\blacksquare$

\section{Predictions of the equivalence class maximum entropy principle}
\label{sec:eq_class_theory}

In this section, we employ the version of the maximum entropy principle (MEP) for quantum state ensembles put forth by Ref.~\cite{chang2025deep}, which we call
``equivalence class MEP'',   to make a prediction of the limiting form of the projected ensemble in the tilted-basis model. It is to be emphasized the equivalence class MEP is only a principle --- it is a useful guide for a first prediction --- but  its validity cannot be taken for granted, and indeed the main result of this work is to identify a robust breakdown of the principle in a class of dynamics.

To begin, consider $N$-qubit RPSs $|\Psi_\pi\rangle$ sampled from the ensemble $\mathcal{F} = \{U_\pi|\Psi_0\rangle:U_\pi\in S_{2^N}\}$.
After performing a projective measurement on subsystem $B$ along the direction $\hat{n}=(\sin\theta_m\cos\phi_m,\sin\theta_m\sin\phi_m,\cos\theta_m)$ for each qubit, the   conditional post-measurement state on subsystem $A$ for measurement outcome $\nu \in \{0,1\}^{N_B}$, averaged over $\mathcal{F}$, is
\begin{equation}
    \bar\rho_A(\nu) = \mathbb{E}_{|\Psi_\pi\rangle\sim\mathcal{F}} \left[ \text{Tr}_B \left( \mathbb{I}_A\otimes (|\Phi_\nu\rangle\langle\Phi_\nu|_B) |\Psi_\pi\rangle\langle\Psi_\pi|  \right) \right]/p(\nu),
\end{equation}
where $p(\nu) = \mathbb{E}_{|\Psi_\pi\rangle\sim\mathcal{F}} \left[ \text{Tr}_{AB} \left( \mathbb{I}_A\otimes (|\Phi_\nu\rangle\langle\Phi_\nu|_B) |\Psi_\pi\rangle\langle\Psi_\pi|  \right) \right]$ denotes the Born probability. Each bit $0(1)$ in the bit-string $\nu$ corresponds to locally obtaining an outcome aligned with (away from) $\hat{n}$.

Following \cite{chang2025deep}, we say that two outcomes $\nu_1$ and $\nu_2$ belong to the same equivalence class, denoted $[\nu]$, if  $\bar{\rho}_A(\nu_1) = \bar{\rho}_A(\nu_2)$. 
Due to the permutation invariance of the ensemble $\mathcal{F}$, all outcomes with the same number of aligned results are equivalent. 
Thus, an equivalence class $[\nu]$ can be uniquely labeled by the integer $\nu_+ = 0,\ldots,N_B$, which counts the number of qubits aligned with $\hat{n}$, i.e., the number of $0$'s appearing in bit-strings $\nu \in [\nu]$. 
The probability of class $[\nu]$ is $p([\nu]) = \sum_{\nu\in[\nu]} p(\nu)$, and we denote the  density matrix associated with class $[\nu]$  as 
\begin{align}
\rho_{[\nu]} : = \bar{\rho}_A(\nu \in [\nu]).
\end{align}

According to the theory, the projected ensemble constructed from a {\it typical} $|\Psi_\pi\rangle \sim \mathcal{F}$, in the limit as $N \to \infty$, consists of a linear combination of pure-state unravelings $\mathcal{E}_{\mathrm{Scrooge}}[\rho_{[\nu]}]$ of the density matrix $\rho_{[\nu]}$. Here $\mathcal{E}_{\mathrm{Scrooge}}[\rho_{[\nu]}]$  is the most entropic ensemble (precisely captured by the minimality of accessible information) whose first moment is   $\rho_{[\nu]}$.  
Thus, the prediction is that the projected ensemble tends to
the generalized Scrooge ensemble (GSE)~\cite{chang2025deep,mark2024maximum}:
\begin{equation}
    \mathcal{E}_\text{PE} 
    \stackrel{N\to\infty}{\rightarrow} \mathcal{E}_{\mathrm{GSE}} = \sum_{[\nu]}p([\nu])\mathcal{E}_{\mathrm{Scrooge}}[\rho_{[\nu]}];
\end{equation}

Our task reduces to calculating $\rho_{[\nu]}$ and $p([\nu])$, forming $\mathcal{E}_{\mathrm{Scrooge}}[\rho_{[\nu]}]$,  and summing them up.

\subsection{Average post-measurement state $\rho_{[\nu]}$}
%Consider an $N$-qubit bipartite system with $N_A$-qubit subsystem $A$ and $N_B$-qubit subsystem $B$.   Starting from a random permutation state  $|\Psi_\pi\rangle = U_\pi|\Psi_0\rangle$ defined in the tilted-basis model,  
% \ww{Chang: in the main text I used $|\Phi_\nu_{\hat{n}}\rangle_B$ as the post-measurement state on subsystem $B$. to explicit capture the $\hat{n}$ dependence. can you make consistent the rest of the sup?}
We  first compute the global  density matrix $\bar\rho := \mathbb{E}_{U_\pi} \left[ U_{\pi} |\Psi_0\rangle\langle\Psi_0| U_{\pi}^\dagger  \right]$, averaged over $\mathcal{F}$. %over the ensemble $\mathcal{F}$,
This can be evaluated using the second-moment operator of the symmetric group:
\begin{equation}
    \mathbb{E}_{U_\pi} [U_{\pi}\otimes U_{\pi}^* ] = \frac{1}{2^N-1}|\sigma_1\rangle\langle\sigma_1|+\frac{1}{2^N(2^N-1)}|\sigma_2\rangle\langle\sigma_2|-\frac{1}{2^N(2^N-1)}|\sigma_1\rangle\langle\sigma_2|-\frac{1}{2^N(2^N-1)}|\sigma_2\rangle\langle\sigma_1|,
\end{equation}
where $\sigma_1$ and $\sigma_2$ denote the two partitions of $Z_2 = \{1,2\}$, and the corresponding  basis states defined in Eq.~\eqref{permutation_basis1}. The overlaps between these basis states and the initial state $|\Psi_0\rangle$ are,
\begin{align}
    \langle \sigma_1 | \Psi_0 \otimes \Psi_0^* \rangle &= 1, \\
    \langle \sigma_2 | \Psi_0 \otimes \Psi_0^* \rangle &= \left[ \left( \cos(\theta_0/2)+ \sin(\theta_0/2) e^{i\phi_0} \right) \left( \cos(\theta_0/2) + \sin(\theta_0/2) e^{-i\phi_0} \right) \right]^N = f^N,
\end{align}
with $f = 1 + \sin\theta_0 \cos\phi_0\in[0,2)$. We exclude the special cases $\{\theta_0=\pi/2,\phi_0=0$\} since the corresponding $|\Psi_0\rangle$ is invariant under $U_\pi$.
We then have 
\begin{equation}
    \bar\rho = \frac{1}{2^N-1} \mathbb{I} +\frac{f^{N}  }{2^N(2^N-1)}M-\frac{f^{N}  }{2^N(2^N-1)} \mathbb{I} -\frac{1  }{2^N(2^N-1)}M,
\end{equation}
where $M$ is a $2^N \times 2^N$ matrix with all entries equal to $1$. %Except in the special case $f = 2$—which corresponds to the initial state being $|x+\rangle^{\otimes N}$—the last two terms are negligible compared to the first two in the large-$N$ limit. 

% We now consider general rank-1 local measurements, where each qubit is measured along the direction $\hat{n}$ on the Bloch sphere. This 
After performing the projective measurement $|\Phi_\nu\rangle\langle\Phi_\nu|_B$ (here we are using similar notation as the main text where $|\Phi_\nu\rangle_B$ is the product state locally pointing along(away from) $\hat{n}$, determined by the measurement outcome $\nu$), we   obtain the
corresponding average post-measurement density matrix of subsystem $A$:
\begin{equation}
    \rho_{[\nu]} =\left( \frac{1-(f/2)^N}{2^N-1}\mathbb{I}_A+\frac{(f^{N}-1) g_+^{\nu_+} g_-^{N_B-\nu_+} }{2^N(2^N-1)}M_A \right)/p(\nu),\label{rho_nu}
\end{equation}
 where the measurement-dependent factors $g_+$ and $g_-$ are 
\begin{equation}
    g_+ = 1+\sin\theta_m\cos\phi_m,\quad g_-=1-\sin\theta_m\cos\phi_m,
\end{equation}
and the Born probability  is 
\begin{equation}
   p(\nu)=  2^{N_A} \frac{(2^N-f^N)+(f^{N}-1) g_+^{\nu_+} g_-^{N_B-\nu_+}}{2^N(2^N-1)}.\label{eq_class_p}
\end{equation}
The probability of the class $[\nu]$ is then
\begin{align}
p([\nu]) = \tbinom{N_B}{\nu_+} p(\nu).
\end{align}

\subsection{GSE equals Haar}
We now show that, for arbitrary local measurement directions $\hat{n}$, the GSE converges in distribution to the Haar ensemble in the thermodynamic limit.
This will be shown in two steps. First, we identify a subset of equivalence classes $[\nu]$, 
distinguished by lying within a ``concentration interval'' to be defined below, whose contributions to the GSE dominate in the thermodynamic limit (i.e., have unity measure).
This allows us to discard the contributions of equivalence classes outside this interval.
Second, we establish that $\rho_{[\nu]}$ for $[\nu]$ in the concentration interval all tend to the maximally mixed state $\mathbb{I}_A/d_A$. The Scrooge ensemble associated with $\mathbb{I}_A/d_A$ is the Haar ensemble, and thus the GSE as a whole is also the Haar ensemble. 

As mentioned, a measurement outcome equivalence class $[\nu]$ may be uniquely labeled by an integer $\nu_+ \in \{0,1,\cdots,N_B\}$ which denotes the number of `0's in the bit-string $\nu$, and hence we will also use $\nu_+$ as  referring to an equivalence class. Let us define the ``concentration interval'' for a given system size $N$ as the interval $[N_B/2 - N_B^\alpha,\, N_B/2 + N_B^\alpha]$, for  any $1/2 < \alpha < 1$.  

% We first establish that, for any $1/2 < \alpha < 1$, the probability that the measurement outcome class $[\nu]$ (labeled by $\nu_+$) lies within the ``concentration interval'' $[N_B/2 - N_B^\alpha,\, N_B/2 + N_B^\alpha]$ approaches unity in the TDL. Classes outside this region thus contribute negligibly to the GSE.

Now from Eq.~\eqref{eq_class_p}, the probability that equivalence class $\nu_+$ falls in the concentration interval is
\begin{align}
     \mathbb{P}(\nu_+\in [N_B/2 - N_B^\alpha,\, N_B/2 + N_B^\alpha] ) &= \sum_{\nu_+\in [N_B/2 - N_B^\alpha,\, N_B/2 + N_B^\alpha]}
    \tbinom{N_B}{\nu_+}2^{N_A} \frac{(2^N-f^N)+(f^{N}-1) g_+^{\nu_+} g_-^{N_B-\nu_+}}{2^N(2^N-1)}.\label{concentration_interval_p}
\end{align}
We analyze two cases: 

(i) When $0\le f \le 1$, we can bound this probability as
\begin{align}
    \mathbb{P}(\nu_+\in [N_B/2 - N_B^\alpha,\, N_B/2 + N_B^\alpha] ) 
    &\geq \frac{1}{(2^N-1)}\sum_{\nu_+\in [N_B/2 - N_B^\alpha,\, N_B/2 + N_B^\alpha]}
    (\frac{\tbinom{N_B}{\nu_+}(2^N-f^N)}{2^{N_B}}-\frac{\tbinom{N_B}{\nu_+}g_+^{\nu_+} g_-^{N_B-\nu_+}}{2^{N_B}})\\
    &\geq \frac{(2^N-f^N)}{(2^N-1)}\sum_{\nu_+\in [N_B/2 - N_B^\alpha,\, N_B/2 + N_B^\alpha]}\frac{\tbinom{N_B}{\nu_+}}{2^{N_B}}-\sum_{\nu_+}\frac{\tbinom{N_B}{\nu_+}g_+^{\nu_+} g_-^{N_B-\nu_+}}{2^{N_B}(2^N-1)}\\
    &\geq \frac{(2^N-f^N)}{(2^N-1)}(1-2e^{-2N^{2\alpha-1}})-\frac{1}{2^N-1},
\end{align}
where the final inequality follows from Hoeffding’s inequality:
\begin{equation}
    \mathbb{P}(|X - N_B/2| \ge N_B^\alpha) \le 2 e^{-2 N_B^{2\alpha - 1}},
    \qquad X \sim \mathrm{Bin}(N_B, 1/2).
\end{equation}

(ii) When $1<f<2$, a similar argument gives
\begin{align}
     \mathbb{P}(\nu_+\in [N_B/2 - N_B^\alpha,\, N_B/2 + N_B^\alpha] ) 
    &\geq \frac{(2^N-f^N)}{(2^N-1)}\sum_{\nu_+\in [N_B/2 - N_B^\alpha,\, N_B/2 + N_B^\alpha]}\frac{\tbinom{N_B}{\nu_+}}{2^{N_B}}\\
    &\geq \frac{(2^N-f^N)}{(2^N-1)}(1-2e^{-2N^{2\alpha-1}}),
\end{align}

In both cases, the right-hand side of Eq.~\eqref{concentration_interval_p} approaches unity as $N \to \infty$.

We therefore ignore equivalence classes with $\nu_+$ outside the concentration interval, and we henceforth restrict our analysis to classes within the concentration region. 
We now compute $\rho_{[\nu]}$. 

From  Eq.~\eqref{rho_nu} , 
we see that  $\rho_{[\nu]}$ is just a linear combination of two operators $\mathbb{I}_A$ and $M_A$. We define the ratio of their coefficients as
\begin{align}
    c([\nu]) = \frac{(f^N-1)\, g_+^{\nu_+}\, g_-^{N_B - \nu_+}}{2^N-f^N} &= (f^N-1)\frac{(1+\sin\theta_m\cos\phi_m)^{\nu_+}(1-\sin\theta_m\cos\phi_m)^{N_B-\nu_+}}{2^N-f^N}\\
    &=\frac{f^N-1}{2^N-f^N} (1-\sin^2\theta_m\cos^2\phi_m)^{N_B/2} (\frac{1+\sin\theta_m\cos\phi_m}{1-\sin\theta_m\cos\phi_m})^{\nu_+-N_B/2}.
\label{eq:ratio}
\end{align}

%-------------------------

Taking the logarithm to analyze the scaling in the thermodynamic limit we get
\begin{equation}
    \log c([\nu]) = \log\left( \frac{f^N - 1}{2^N - f^N} \right) + N_B/2 \log(1-\sin^2\theta_m \cos^2\phi_m) + (\nu_+ - N_B/2) \log\left( \frac{1 + \sin\theta_m\cos\phi_m}{1 - \sin\theta_m\cos\phi_m} \right).
\end{equation}

For $(\theta_0,\phi_0)\notin \{(\pi/2,0)\}$, we have $f < 2$, so $f^N/2^N \to 0$ as $N \to \infty$, making the first term negative and extensive. 
Within the concentration interval $\nu_+ \in [N_B/2 - N_B^\alpha,\, N_B/2 + N_B^\alpha]$ with $1/2 < \alpha < 1$, the last term grows at most $\mathcal{O}(N_B^\alpha)$, which is subleading compared to the extensive negative second term.  
Hence, for all typical $\nu_+$, % the dominant term is negative and extensive in $N_B$, implying $\log c([\nu]) \to -\infty$ as $N \to \infty$. Hence, we conclude that
\begin{equation}
    \log c([\nu]) \to -\infty \quad \Rightarrow \quad c([\nu]) \to 0 \quad \text{as } N \to \infty.
\end{equation}

%for all  $\nu_+$ in the concentration interval.

This shows that the $M_A$ component in $\rho_{[\nu]}$ vanishes in the thermodynamic limit, so
\begin{equation}
    \lim_{N\to\infty} \rho_{[\nu]} = \frac{\mathbb{I}_A}{d_A},
\end{equation}
for all typical $[\nu]$ in the concentration interval. Note the complete $\hat{n}$ independence of this result.

Thus, according to the equivalence class MEP,
for a typical $|\Psi_\pi\rangle \sim \mathcal{F}$, excluding $\theta_0  = \pi/2$ and $\phi_0=0$, we have the prediction
\begin{align}
\mathcal{E}_{\text{PE}} \stackrel{\text{MEP}}{\rightarrow} \mathcal{E}_\text{GSE} \stackrel{N \to \infty}{\rightarrow} \mathcal{E}_\text{Haar},
\end{align}
independent of $\hat{n}$. 
%the limiting form of the PE
%
% Consequently, according to the equivalence-class maximum-entropy principle, the limiting form of the PE (i.e., the GSE) converges to the Haar ensemble in the thermodynamic limit for arbitrary $\hat{n}$. However, this prediction is in contrast with our results of the classical bit-string ensemble. Hence, our model provides an explicit counterexample to the maximum entropy principle.

\section{Projected ensemble of
the tilted-basis model: analytical results}
\label{sec:PE_RPS_analytic}

% \ww{I have not gone through below yet}

% In the main text we study the projected ensemble  generated from a random permutation state by performing projective measurements on subsystem $B$. Our central result of the tilted-basis model is that the limiting form of the PE undergoes a phase transition: as the measurement basis is varied from parallel to perpendicular to the computational ($z$) basis, the limiting ensemble changes from a minimally ergodic classical bit-string ensemble to a maximally ergodic Haar ensemble. 

In this section, we provide details of the form of the projected ensemble   within the tilted-basis model, in the special scenarios of $x$-basis and $z$-basis measurements. The statement made in the main text is that these are the Haar and classical bit-string ensembles respectively. 
% We also provide details on numerical simulations performed to ascertain the deep thermalization phase transition presented in Fig.~1(b) of the main text. 
%
%the two limiting measurement directions (the $x$-basis and $z$-basis measurements), together with  numerics for the PE phase transition. 
Throughout this section, we focus on the  random permutation states $|\Psi_\pi\rangle$ of the tilted-basis model defined in Eq.~\eqref{RPS}, with parameters $(\theta_0,\phi_0) \notin \{\theta_0$\,$=$\,$0,\theta_0$\,$=$\,$\pi,(\theta_0$\,$=$\,$\tfrac{\pi}{2},\phi_0$\,$=$\,$0)\}$, such that the subsystem $A$ always thermalizes by Theorem 1.

%def of PE , general measuerment 
%besidies, we also show the prediction of the equivalence class theory.

\subsection{$x$-basis measurement: (average) projected ensemble equals the Haar ensemble}

For $x$-basis measurements considered here, we denote measurement outcomes by $x_B \in \{0,1\}^{N_B}$. 
%
% We first consider constructing the projected ensemble by measuring subsystem $B$ in the $x$-basis. Performing a projective measurement on $B$ in the complete basis $\{|x_B\rangle\}$ yields a conditional pure state on $A$ for each measurement outcome $x_B$:
% \begin{equation}
%    |\psi_A(x_B)\rangle = \frac{1}{\sqrt{p(x_B)}} (\mathbb{I}_A \otimes \langle x_B|) |\Psi_{\pi}\rangle,
%\end{equation}
%where $p(x_B) = \langle \Psi_{\pi}| (\mathbb{I}_A \otimes |x_B\rangle\langle x_B|) |\Psi_{\pi}\rangle$ is the Born probability of observing outcome $x_B$.
%
%This defines the projected ensemble:
%\begin{equation}
%    \mathcal{E}_{\mathrm{PE}} := \{p(x_B), |\psi_A(x_B)\rangle\}.
%\end{equation}
%
Our aim here to show that the $k$-th moment of the projected ensemble, averaged over generator states $|\Psi_\pi \rangle$, with $\pi$ sampled uniformly from the symmetric group $S_{2^N}$, matches that of the Haar ensemble in the thermodynamic limit, i.e., 
\begin{equation}
    \mathbb{E}_{U_\pi}[\rho^{(k)}_{\mathrm{PE}}]\;\stackrel{N \to \infty}{\to}\;\rho^{(k)}_{\mathrm{Haar}},\label{PEtoHaar}
\end{equation}
where $\mathbb{E}_{U_\pi}[\cdot]$ denotes the expectation over permutations.  This provides a non-trivial consistency check  that a {\it typical} $\rho^{(k)}_\text{PE}$ approaches $\rho^{(k)}_\text{Haar}$.

%------------------------------
\subsubsection{$k$-th moment of the unnormalized projected states }

However,  the $k$-th moment is a rational function of permutation unitaries,
\begin{align}
    \mathbb{E}_{U_\pi}[\rho^{(k)}_{\mathrm{PE}}] &= \mathbb{E}_{U_\pi} \left[\sum_{x_B} p(x_B)(|\psi_A(x_B)\rangle\langle\psi_A(x_B)|)^{\otimes k}\right]\\
    &=\sum_{x_B}\mathbb{E}_{U_\pi} \left[
    \frac{((\mathbb{I}_A\otimes \langle x_B|)U_\pi|\Psi_0\rangle\langle \Psi_0|U_\pi^\dagger (\mathbb{I}_A\otimes|x_B\rangle))^{\otimes k}}{(\langle \Psi_0|U_\pi^\dagger (\mathbb{I}_A\otimes|x_B\rangle)(\mathbb{I}_A\otimes \langle x_B|)U_\pi|\Psi_0\rangle)^{ k-1}}\right],
\end{align}
which cannot be handled directly by Weingarten calculus. 
To circumvent this, we instead analyze the  unnormalized post-measurement state, 
\begin{equation}
    |\tilde\psi_A(x_B)\rangle := (\mathbb{I}_A \otimes \langle x_B|)\,|\Psi_\pi\rangle,
\end{equation}
and its statistical moments $ \mathbb{E}_{U_\pi}\left[|\tilde\psi_A(x_B)\rangle\langle\tilde{\psi}_A(x_B)|^{\otimes k} \right] $.

We  claim %\textcolor{blue}{(Eq.~\eqref{...} in the main text)} 
that for a fixed measurement outcome $x_B\neq(+,\ldots,+)$ and parameters $(\theta_0,\phi_0)\notin\{0,\pi\}$, the averaged $k$-th moment of the unnormalized projected state coincides with that of the so-called Ginibre ensemble in the TDL:
\begin{equation}
\mathbb{E}_{U_\pi}\left[|\tilde\psi_A(x_B)\rangle\langle\tilde\psi_A(x_B)|^{\otimes k}\right]
    = \frac{1}{d^k}\sum_{\tau\in S_k}\text{Perm}_{\mathcal{H}_A^{\otimes k}}(\tau) + o\left(\frac{1}{d^k}\right).
    \label{supp_komentsHaar}
\end{equation}
 % with subsystem Hilbert space dimension $d_A$ fixed while $d\to\infty$.

To establish this, we use the Haar integral over the symmetric group in Eq.~\eqref{Haar_permutation},  which gives the averaged $k$-th moment in the form
\begin{equation}
\mathbb{E}_{U_\pi}\left[|\tilde\psi_A(x_B)\rangle\langle\tilde{\psi}_A(x_B)|^{\otimes k} \right] = \sum_{i,j}\text{Wg}^{(2k)}(\sigma_i,\sigma_j)  |\sigma_i\rangle_A \langle x_B^{\otimes 2k}|\sigma_i\rangle_B\langle\sigma_j|\Psi_0^{\otimes k}\otimes\Psi_0^{*\otimes k}\rangle,
\end{equation}
where $\text{Wg}^{(2k)}$ is the $2k$-th order Weingarten function, $\sigma_i$ denotes a partition of the set $Z_{2k} = \{1,2,\dots,2k\}$, $|\sigma_i\rangle$ is the corresponding global basis state, and $|\sigma_i\rangle_{A,B}$ are the restrictions of this basis to subsystems $A$ and $B$.
To determine the scaling of the coefficient  $(\text{Wg}^{(2k)}(\sigma_i,\sigma_j)\langle x_B^{\otimes 2k}|\sigma_i\rangle_B\langle\sigma_j|\Psi_0^{\otimes k}\otimes\Psi_0^{*\otimes k}\rangle)$ in the large $d$ limit, we  analyze the scaling of inner product $\langle x_B^{\otimes 2k}|\sigma_i\rangle_B$, as follows. %and $\langle\sigma_j|\Psi_0^{\otimes k}\otimes\Psi_0^{*\otimes k}\rangle$ separately.

Consider a partition $\sigma=\{D_1,D_2,\cdots,D_r\}$, where $D_i$ represents a block of size $|D_i|$. We have:
\begin{equation}
    \langle x_B^{\otimes 2k}|\sigma\rangle_B = \prod_{i=1}^r\langle x_B^{\otimes |D_i|}|D_i\rangle_B.
\end{equation}
Next consider a single block $D \in \sigma$ with size $|D|$. In this case,  
\begin{equation}  
\langle x_B^{\otimes |D|}|D\rangle_B =\prod_{l=1}^{N_B}
    \left(\sum_{a=1}^2\langle x_{B,l}^{\otimes |D|}|a^{\otimes |D|}\rangle\right) = \left\{  
     \begin{array}{lr}  
      0,\quad  |D|=2p+1&  \\  
   1/d_B^{p-1},\quad |D|=2p &    
     \end{array}.  
\right.  
\end{equation} 
Since we are considering $x_B \neq (+,\dots,+)$, whenever $|D|$ is odd there must exist at least one site with measurement outcome $|x_{B,l}\rangle = |-\rangle$, which is orthogonal to $\sum_a |a\rangle$. Thus, if any block $D \in \sigma$ has an odd number of elements, we obtain  
$\langle x_B^{\otimes 2k} | \sigma \rangle_B = 0$. Therefore, we have:
\begin{equation}  
\langle x_B^{\otimes 2k}|\sigma\rangle_B  = \left\{  
     \begin{array}{lr}  
      0,\quad  \text{if\quad} |D_i|=2p+1,&  \\  
   d_B^{r-k}, &    
     \end{array}  
\right.  
\end{equation} 
where $r=\#\sigma$.

Then we analyze the inner product with the initial state $\Psi_0$,
\begin{equation}
    \langle\sigma|\Psi_0^{\otimes k}\otimes\Psi_0^{*\otimes k}\rangle = \prod_{i=1}^r\langle D_i|\Psi_0^{\otimes n_{D_i}}\otimes\Psi_0^{*\otimes (|D_i|-n_{D_i}) }\rangle,
\end{equation}
where $n_{D_i}$ and $|D_i|-n_{D_i}$ denote the number of $\Psi_0,\Psi_0^*$ in the block $D_i$. For each block $D$, we have:
\begin{equation}
    w(D,\Psi_0) := |\langle D|\Psi_0^{\otimes n_D}\otimes\Psi_0^{*\otimes (|D|-n_D) }\rangle|
    =|\cos(\theta_0/2)^{|D|}+\sin(\theta_0/2)^{|D|}\exp(i(|D|-2n_D)\phi_0)|^N,
\end{equation}
Therefore, we obtain
\begin{equation}  
w(D,\Psi_0)  = \left\{  
     \begin{array}{lr}  
     o(d^{1/2}), \quad \text{if\quad} |D|=1,&\\
      1,\quad  \text{if\quad} |D|=2, \quad\text{and}\quad n_D=1,&  \\  
      o(1),\quad \text{if\quad} |D|>2.&
     \end{array}  \label{wDpsi}
\right.  
\end{equation} 
  Note that only when $|D|=2$, and $n_D=1$, we have $w(D,\Psi_0)=1$. If $|D|=2, n_D\neq1$, then $w(D,\Psi_0) = |\cos^2(\theta_0/2)+\sin^2(\theta_0/2)e^{i2\phi_0}|^N\ll 1$ for $\phi_0\neq 0,\pi $. 

%\subsubsection{Proof of Eq.~\eqref{kmoments}}
We next consider the asymptotic form of the Weingarten function in Eq.~\eqref{asymptotic_Wg}. 
Since the M{\"o}bius function is independent of $d$, we have $\text{Wg}^{(2k)}(\sigma_i,\sigma_j)\sim d^{-\#(\sigma_i\wedge\sigma_j)}$ in the large-$d$ limit.

%To analyze the scaling behavior of the $k$-th moment  $ \mathbb{E}_{U_\pi}\left[|\tilde\psi_A(x_B)\rangle\langle\tilde{\psi}_A(x_B)|^{\otimes k} \right] $,we further introduce the function
Combining this with the previous analysis, we introduce the function
\begin{align}
    f(\sigma_i,\sigma_j,\Psi_0) := |d^{-\#(\sigma_i\wedge\sigma_j)}\langle x_B^{\otimes 2k}|\sigma_i\rangle_B\langle\sigma_j|\Psi_0^{\otimes k}\otimes\Psi_0^{*\otimes k}\rangle|
\end{align}
to characterize the scaling behavior of the $k$-th moment $ \mathbb{E}_{U_\pi}\left[|\tilde\psi_A(x_B)\rangle\langle\tilde{\psi}_A(x_B)|^{\otimes k} \right] $.

We restrict to the case where all blocks of $\sigma_i$ contain an even number of elements, 
since the presence of any odd block in $\sigma_i$ yields $f=0$. Then, $f(\sigma_i,\sigma_j,\Psi_0)$ can be expressed as
\begin{equation}
    f(\sigma_i,\sigma_j,\Psi_0)=d_A^{k-r_i} d^{-t+r_i-k}\prod_{m=1}^{r_j}w(D_m,\Psi_0),
\end{equation}
where $r_i=\#\sigma_i, r_j=\#\sigma_j$, and $t = \#(\sigma_i\wedge\sigma_j)$, and $D_m$ denotes a block of the partition $\sigma_j$.

Using the bounds on $w(D,\Psi_0)$ from Eq.~\eqref{wDpsi}, we have
\begin{equation}
    \prod_{m=1}^{r_j}w(D_m,\Psi_0)\lesssim d^{\,s_1/2},
\end{equation}
where $s_1$ denotes the number of singletons ($|D|=1$ blocks) in $\sigma_j$.
Each singleton contributes at most $o(d^{1/2})$, each 2-block contributes at most $1$, and larger blocks contribute additional $o(1)$ factors. Therefore, we can bound
\begin{equation}
    f(\sigma_i,\sigma_j,\Psi_0) \lesssim d^{\,r_i-k + s_1/2 - t}.
\end{equation}
 We then  show that $r_i + s_1/2 - t \le 0$, with strict inequality whenever $\sigma_i$ and $\sigma_j$ are not both pairings.

Fix a block $D\in\sigma_i$ of size $2l$. Let $s(D)$ be the number of $\sigma_j$-singletons contained in $D$. The number of distinct $\sigma_j$-blocks that meet $D$ is at least $s(D)$, and if $D$ is not entirely composed of singletons it is at least $s(D)+1$. Summing over all $\sigma_i$-blocks gives
\begin{equation}
    t \geq \sum_D s(D)+\#\{\text{\(\sigma_i\)-blocks that contain a non-singleton}\}.
\end{equation}
Since $\sum_D s(D)=s_1$, and if $u$ denotes the number of $\sigma_i$-blocks made entirely of $\sigma_j$-singletons then the second term equals $r_i-u$. Thus
\begin{equation}
t \ge s_1 + (r_i-u) = r_i + s_1 - u.
\end{equation}
Each $\sigma_i$-block made entirely of singletons has size at least $2$, so $u\le s_1/2$. Hence
\begin{equation}
t \ge r_i + s_1 - u \ge r_i + \frac{s_1}{2},    
\end{equation}

i.e., $r_i + s_1/2 - t \le 0$. Consequently we always have $f(\sigma_i,\sigma_j,\Psi_0)\lesssim d^{-k}$.

For $f(\sigma_i,\sigma_j,\Psi_0)$ to be of order $d^{-k}$ we need equality at every inequality above. In particular:
\begin{itemize}
  \item No $o(1)$ loss from $w(D,\Psi_0)$: thus $\sigma_j$ has no blocks of size $>2$.
  \item No $o(d^{1/2})$ loss: hence $s_1=0$ (no singletons).%$\ll d^{1/2}$ 
  \item Equality $t=r_i$: every $\sigma_i$-block meets exactly one $\sigma_j$-block, so $\sigma_i$ refines $\sigma_j$.
\end{itemize}
With $s_1=0$ both partitions are pairings (each has $k$ blocks of size 2). The refinement then forces $\sigma_i=\sigma_j$. For the initial state $\Psi_0$ with parameter $\phi_0\neq0,\pi$, only the class with $|D|=2$ and $n_D=1$ satisfies $w(D,\Psi_0)=1$.
%only $|D|=2,n_D=1$ we have $w(D,\Psi_0)=1$. 
Thus, these pairings are permutations in the $k$-replica space, i.e. $\sigma_i=\sigma_j=\tau_i\in S_k$. By considering the M{\"o}bius function,
\begin{equation}
    \mu(\tau,\tau)=1.
\end{equation}
Finally, we arrive at Eq.~\eqref{supp_komentsHaar}, which establishes our statement of the un-normalized projected state.

%\textcolor{blue}{I do not show here, but when $\phi = 0,\pi$, we recover the orthogonal Haar case, since for any pairing with $|D| = 2$, we  have $w(D,\Psi_0) = 1$.
%Although orthogonality is absent, analogous results can be obtained by considering measurements in the $X$-$Y$ plane.}

\subsubsection{Replica trick}

Now,  we use information about the {$k'$-th} moments of the average unnormalized projected state to reveal information about the {$k$-th} moment of the average normalized projected state, through the replica trick.
% We will invoke an analytic continuation of the former to 

We introduce 
%\begin{align}
 %   \rho^{(n,k)}_{\mathrm{PE}} &= \sum_{x_B}\frac{\mathbb{E}_{U_\pi} [\langle\tilde\psi_A(x_B)|\tilde\psi_A(x_B)\rangle^{n}(|\tilde\psi_A(x_B)\rangle\langle\tilde\psi_A(x_B)|)^{\otimes k}]}{\mathbb{E}_{U_\pi} [\langle\tilde\psi_A(x_B)|\tilde\psi_A(x_B)\rangle^{ n+k-1}]},\label{rho_nk}
%\end{align}
the intermediary object 
\begin{align}
    \rho^{(n,k)}_{\mathrm{PE}} &= \sum_{x_B}\mathbb{E}_{U_\pi} [\langle\tilde\psi_A(x_B)|\tilde\psi_A(x_B)\rangle^{n}(|\tilde\psi_A(x_B)\rangle\langle\tilde\psi_A(x_B)|)^{\otimes k}],\label{rho_nk}
\end{align}
which upon taking the limit $n\to 1-k$  will recover
\begin{equation}
    \lim_{n\to1-k}\rho^{(n,k)}_{\mathrm{PE}} = \mathbb{E}_{U_\pi}[\rho^{(k)}_{\mathrm{PE}}],
\end{equation}
where  
\begin{align}
    \mathbb{E}_{U_\pi}\left[\rho^{(k)}_{\mathrm{PE}} \right] &= \sum_{x_B}\mathbb{E}_{U_\pi} \left[\frac{(|\tilde\psi_A(x_B)\rangle\langle\tilde\psi_A(x_B)|)^{\otimes k}}{\langle\tilde\psi_A(x_B)|\tilde\psi_A(x_B)\rangle^{ k-1}} \right].
\end{align}
We observe that $\rho_\text{PE}^{(n,k)}$ is analytically computable for integer $k$ and integer $n$ (directly from our results in the previous subsection) as there it is a rational function of unnormalized states; thus the step of taking the limit $n \to 1-k$ from such discrete integer data is an analytic continuation of $\rho_\text{PE}^{(n,k)}$ to the real plane. This procedure of taking the `replica limit' may not be fully mathematically rigorous, but is often used in physics e.g., in spin-glass computations~\cite{edwards1975theory}.

% Using the conclusion for the unnormalized state in
Using Eq.~\eqref{supp_komentsHaar}, we find that for each measurement outcome $x_B \neq (+,\ldots,+)$,
\begin{align}
    &\mathbb{E}_{U_\pi} [\langle\tilde\psi_A(x_B)|\tilde\psi_A(x_B)\rangle^{n}(|\tilde\psi_A(x_B)\rangle\langle\tilde\psi_A(x_B)|)^{\otimes k}]\\ =& \text{Tr}_{\mathcal{H}_A^{\otimes n}}( \mathbb{E}_{U_\pi}[(|\tilde\psi_A(x_B)\rangle\langle\tilde\psi_A(x_B)|)^{\otimes n+k}])\\
    =&\frac{1}{d^{n+k}}\sum_{\tau\in S_{n+k}}\text{Tr}_{\mathcal{H}_A^{\otimes n}}(\text{Perm}_{\mathcal{H}_A^{\otimes {n+k}}}(\tau)) + o\left(\frac{1}{d^{n+k}}\right)\\
    =&  \frac{\prod_{j=0}^{n-1}(d_A+k+j)}{d^{n+k}}\sum_{\tau\in S_k}\text{Perm}_{\mathcal{H}_A^{\otimes k}}(\tau) + o\left(\frac{1}{d^{n+k}}\right)\\
    =&\frac{\Gamma(d_A+k+n)}{\Gamma(d_A+k)d^{n+k}}\text{Perm}_{\mathcal{H}_A^{\otimes k}}(\tau) + o\left(\frac{1}{d^{n+k}}\right).
\end{align}
Notice that in the last equality, we wrote the factorial (a product of integers) in terms of the continuous Gamma function $\Gamma(x)$; this is the `analytical continuation' step.
Summing over $x_B$ and taking the limit $n\to 1-k$, we obtain
\begin{equation}
    \lim_{n\to1-k}\rho^{(n,k)}_{\mathrm{PE}} = \frac{1}{d_A(d_A+1)\cdots(d_A+k-1)}\sum_{\tau\in S_k}\text{Perm}_{\mathcal{H}_A^{\otimes k}}(\tau)+o(1)  = \rho_{\mathrm{Haar}}^{(k)}+o(1).
\end{equation}
Finally, assuming we discard the contribution from the outcome $x_B=(+,\cdots,+)$, whose probability  should vanish in the thermodynamic limit, we arrive at our claim of the average PE generated by the $x$-basis measurements:
\begin{equation}
     \mathbb{E}_{U_\pi}[\rho^{(k)}_{\mathrm{PE}}]\;\ \stackrel{N\to\infty}{\to}\;\rho^{(k)}_{\mathrm{Haar}}.
\end{equation}

% Ideally, we would like to also make the claim that a {\it typical} $\rho^{(k)}_{\mathrm{PE}}$ (and not jut the average)  approaches $\rho^{(k)}_{\mathrm{Haar}}$ with high probability too, but we expect from concentration of measure statements that this will occur. 

\subsection{$z$-basis measurement:  emergence of the classical bit-string ensemble}
 
We now turn to the complementary limiting case of $z$-basis measurements. We denote measurement outcomes by $z_B \in \{0,1\}^{N_B}$. 
Our   claim  in the main text is that the projected ensemble for a typical RPS tends to the classical bit-string ensemble
\begin{equation}
     \mathcal{E}_{\mathrm{PE}} \stackrel{N \to \infty}{\to} \mathcal{E}_{\mathrm{Cl}},\quad\mathcal{E}_{\text{Cl}} := \left\{p(z_A) = 1/d_A,\, |z_A\rangle\right\},
\end{equation}
an ensemble of pure states uniformly distributed over the computational basis states $|z_A\rangle$ on $A$. Here $d_A$ is the dimension of the Hilbert space $\mathcal{H}_A$. 

Instead of showing this, we will be content with showing the following simpler statement, which provides a nontrivial test for  such emergence: that the typical projected state $|\psi_A(z_B)\rangle$, randomly generated over RPS but for a fixed $z_B$, tends to some computational basis state $|z_A\rangle$ on $A$. We expect more generally that (i) there is no preference over which $|z_A\rangle$ the projected state of a different measurement outcome $z_B$ tends to; thus, the coverage over $\{|z_A\rangle\}$ should be uniform. (ii) That the projected ensemble for a {\it typical} RPS behaves the same way too.

Concretely, we show:

{\bf Theorem 2.} {\it
Let $|\psi_A(z_B)\rangle$ be an $N_A$-qubit projected state obtained by performing a computational basis measurement on subsystem $B$ of an $N$-qubit random permutation state $|\Psi_{\pi}\rangle$, with measurement outcome $z_B$. %Without loss of generality, assume $0 < \theta < \pi/2$. 
Consider its computational basis decomposition
$|\psi_A(z_B)\rangle = \sum_{z_A} c_{z_A}(z_B) |z_A\rangle,$
and denote the order statistics of the coefficients by
$|c_{(1)}| \geq |c_{(2)}| \geq \cdots \geq |c_{(m)}|.$
Then, for any $\alpha > 0$ and $\theta_0\notin \{0,\pi/2,\pi\}$, fixing $z_B$ and considering projected states generated from RPS, we have
\begin{equation}
    \mathbb{P}\left( \frac{|c_{(1)}|^2}{|c_{(2)}|^2} \geq \omega^{N^\alpha} \right)
    \geq 1 - (C\frac{[N^\alpha]}{\sqrt{N}}+\frac{D}{2^N}),
    %(\frac{2^{N_A}(2^{N_A}-1)}{2} \cdot \frac{2[N^\alpha] + 3}{\sqrt{\pi N}}+ \frac{2^{N_A}(2^{N_A}-1)}{2^{N+1}}),
\end{equation}
where $\omega = \mathrm{max}(\cot^2(\theta_0/2),\tan^2(\theta_0/2)) > 1$, $[x]$ denotes the integer part of $x$, and
$C,D>0$ are constants depend on $N_A$. %[N^\alpha]
}

Theorem 2 just says that the leading coefficient $c_{(1)}$ dominates in the thermodynamic limit, i.e., $|c_{(1)}|\to1$, thus $|\psi_A(z_B)\rangle \to |z_A\rangle$ for some $|z_A\rangle$. 

 We first give a quick sketch of the proof presented below. Consider the computational basis decomposition
$|\psi_A(z_B)\rangle = \sum_{z_A} c_{z_A}(z_B) |z_A\rangle$.  The coefficients $|c_{z_A}(z_B)|$ depend only on the Hamming weight of the corresponding unpermuted bit string $\pi^{-1}((z_A,z_B))$.
(i) We show that the distribution of Hamming weights of the $d_A$ bit strings compatible with fixed $z_B$ is close to the distribution of $d_A$ i.i.d.\ $\mathrm{Binomial}(N,\tfrac12)$ variables;  
(ii) we then bound the separation between the largest and second-largest samples (the top–two gap) in this i.i.d.\ binomial ensemble;  
(iii) we finally transfer this bound to the original coefficients, establishing the desired ratio $|c_{(1)}|^2/|c_{(2)}|^2$.

%Theorem 2 states that, with unit probability in the thermodynamic limit, one of the coefficients $c_{z'_A}$ of the projected state $|\psi_A(z_B)\rangle$ becomes dominant—its magnitude approaches 1 while all others vanish. As a result, in the TDL $N_A \ll N$, the projected state under a random permutation $U_\pi$ converges in probability to a computational basis state $|z'_A\rangle$.% This establishes that the projected ensemble $\mathcal{E}_{PE}^z$ converges to the classical bit-string ensemble $\mathcal{E}_{cl}$ composed of bit-string states.

 \subsubsection{Proof of theorem~2}

\textit{Proof.} Without loss of generality, we restrict to $\theta_0\in(\pi/2,\pi)$, so that $\sin(\theta_0/2)>\cos(\theta_0/2)$ and hence $\omega=\tan^2(\theta_0/2)>1$.

Consider a projective measurement on the full system, with outcome bit string $(z_A,z_B)$. The permutation $U_\pi$ maps each computational basis state $|z\rangle$ to another basis state $|\pi(z)\rangle$, so the Born probability of observing $\pi(z)$ in the state $|\Psi_\pi\rangle$ equals the probability of observing $z$ in the initial state $|\Psi_0\rangle$. Since this probability depends only on the Hamming weight $h(z)$ (the number of ones in bit string $z$), we have
%To analyze the behavior of the coefficient $|c_{z_A}(z_B)|^2$, we consider performing a projective measurement on the entire system and partitioning the measurement outcome into two parts, $(z_A, z_B)$. Since the permutation unitary $U_{\pi}$ maps each computational basis state $|z\rangle$  to another basis state $|\pi(z)\rangle$, the Born probability of obtaining $\pi(z)$ in the permuted state $|\psi_\pi\rangle$ is equal to the Born probability $p_{\text{ini}}(z)$ of obtaining outcome $z$ in the initial state $|\Psi_0\rangle$. Note that $p_{\text{ini}}(z)$ depends only on the Hamming weight $h(z)$, i.e., the number of '1's in the bit string $z$.  Then the Born probability is given by
\begin{equation}
    p(\pi(z)) = p_{ini}(z) = \sin^{2h(z)}(\theta_0/2) \cos^{2(N - h(z))}(\theta_0/2).
\end{equation}
Writing $\pi(z)=(z_A,z_B)$ with $z_A$ on subsystem $A$ and $z_B$ on subsystem $B$, the coefficient $|c_{z_A}(z_B)|^2$ of the projected state $|\psi_A(z_B)\rangle$ is proportional to $p(z_A,z_B)$. Thus, for two bit strings $z_A,z_A'$ associated with the same $z_B$, the ratio of coefficients is
\begin{equation}
    \frac{|c_{z_A}(z_B)|^2}{|c_{z_A'}(z_B)|^2} = \frac{p_{ini}(z)}{p_{ini}(z')} = \tan^{2(h(z)-h(z'))}(\theta_0/2),
\end{equation}
where $\pi(z)=(z_A,z_B)$ and $\pi(z')=(z_A',z_B)$. It therefore suffices to analyze the distribution of the Hamming weight of the full bit string $z$ of the initial state $|\Psi_0\rangle$.

Let
\[
Q\;=\;\text{joint distribution of the Hamming‐weight vector }
\bigl(h(z_1),\dots,h(z_{d_A})\bigr),
\]
where each $z_i$ is a bit string sampled by applying a random permutation $\pi$ under the constraint that $\{\pi(z_1),\cdots,\pi(z_{d_A})\} = \{(z_A,z_B)\}$  with $z_B$ fixed on subsystem $B$.

For comparison, define
\[
Q_B\;=\;\bigotimes_{i=1}^{d_A}\!B\bigl(N,1/2)
\]
the distribution of $d_A$ independent $\mathrm{Binomial}(N,1/2)$ random variables. This distribution can be understood as randomly choosing $d_A$ bit-strings from all $d$ bit-strings with replacement.

Then, we show that the distribution $Q$ is close to the distribution $Q_B$ by computing the total variation distance (TVD).

%Since sampling without replacement from a set of size $2^N$ isequivalent to conditioning $d_A$ independent draws on all being distinct
Firstly, observe that the permutation $\pi \in S_{2^N}$ is chosen uniformly at random. Therefore, the conditional event
$\{\{\pi(z_1), \ldots, \pi(z_{d_A})\} = \{(z_A, z_B)\}\}$
is equivalent to selecting $z_1, \ldots, z_{d_A}$ uniformly at random (without replacement) from all bitstrings on the total system, under the sole constraint that $z_i \neq z_j$ for all $i \neq j$. Thus, we find,
\begin{equation}
    Q = Q_B|\{z_i \,\, \text{all distinct}\}.
\end{equation}
We can compute the TVD between $Q$ and $Q_B$,
\begin{equation}
    \delta_{TVD}(Q,Q_B) = 1-\mathbb{P}(\{z_i  \,\, \text{all distinct}\})=1-\Pi_{j=0}^{d_A-1}(1-\frac{j}{2^N}).
\end{equation}
Using the inequality $1-\Pi(1-a_j)\leq\sum a_j$ for $0<a_j<1$, we can bound the TVD,
\begin{equation}
    \delta_{TVD}(Q,Q_B) \leq \frac{2^{N_A}(2^{N_A}-1)}{2^{N+1}}.
\end{equation}

Therefore, we consider the order statistics of Hamming weight $h(z_i)$ as $h_{(1)}\geq h_{(2)}\geq\cdots\geq h_{(m)} $. For any $\alpha > 0$, let $A = \{|h_{(1)}-h_{(2)}|\leq N^\alpha\}$. Then the total variation bound gives
\begin{equation}
    |Q(A)-Q_B(A)|\leq\delta_{TVD}(Q,Q_B) , \label{QAQB}
\end{equation}

Next, we introduce the following lemma, which provides a bound on the probability that the two largest order statistics of $d_A$ independent $\mathrm{B}(N,1/2)$ random variables differ by at most $N^\alpha$. This lemma is key to establishing Theorem 2.

{\bf Lemma 1.} {\it Let $X_1,X_2,\cdots,X_m$ be independent random variables, each distributed according to the binomial distribution $B(N,1/2)$. Denote their order statistics by $X_{(1)}\geq X_{(2)}\geq\cdots\geq X_{(m)} $. Then, for any $\alpha>0$, we have
\begin{equation}
    \mathbb{P}(|X_{(1)}-X_{(2)}|\leq N^\alpha)
    \leq\frac{m(m-1)}{2}\frac{(2[N^\alpha]+3)}{\sqrt{\pi N}}.
\end{equation}
}

Combining Eq.~\eqref{QAQB} with Lemma 1 yields an explicit upper bound on the probability under the random‐permutation distribution $Q$:
\begin{equation}
    Q(A) \leq Q_B(A)+\delta_{TVD}(Q,Q_B)\leq \frac{2^{N_A}(2^{N_A}-1)}{2} \cdot \frac{2[N^\alpha] + 3}{\sqrt{\pi N}}+ \frac{2^{N_A}(2^{N_A}-1)}{2^{N+1}}.
\end{equation}

Therefore, let $|\psi_A(z_B)\rangle$ be the projected state with computational‐basis expansion $|\psi_A(z_B)\rangle = \sum_{z_A} c_{z_A}(z_B) |z_A\rangle,$and let the magnitudes of its coefficients be ordered as
\[
|c_{(1)}|\ge |c_{(2)}|\ge \cdots \ge |c_{(d_A)}|,
\]
where $d_A=2^{N_A}$.  Then for any $\alpha>0$, setting $A=\{|h_{(1)}-h_{(2)}|\le N^\alpha\}$ and using the bound above gives
\begin{align}
    \mathbb{P}\left( \frac{|c_{(1)}|^2}{|c_{(2)}|^2}  \geq \omega^{N^\alpha} \right) &=\mathbb{P}\left( \frac{\sin^{2h_{(1)}(z)}(\theta_0/2) \cos^{2(N - h_{(1)}(z))}(\theta_0/2)}{\sin^{2h_{(2)}(z)}(\theta_0/2) \cos^{2(N - h_{(2)}(z))}(\theta_0/2)}\geq (\frac{\sin^2\theta_0}{\cos^2\theta_0})^{N^\alpha}
    \right)\\
    &= \mathbb{P}\left(|h_{(1)}-h_{(2)}|\geq N^\alpha \right)
    =Q(\bar A)\\
    &\geq 1 - (\frac{2^{N_A}(2^{N_A}-1)}{2} \cdot \frac{2[N^\alpha] + 3}{\sqrt{\pi N}}+ \frac{2^{N_A}(2^{N_A}-1)}{2^{N+1}}).
\end{align}
For the complementary case $\theta_0\in(0,\pi/2)$, we have $\cos(\tfrac{\theta_0}{2})>\sin(\tfrac{\theta_0}{2})$, and we set $\omega=\cot^2(\tfrac{\theta_0}{2})>1$. The proof proceeds analogously.  
This completes the proof of Theorem 2.$\blacksquare$

\subsubsection{Proof of Lemma 1}
%{\it Proof of Lemma 1.} \\
We first consider the $m=2$ case.
Since $X_2\sim B(N,1/2)$, we have $N-X_2\sim B(N,1/2)$. Then we find,
\begin{equation}
    \mathbb{P}(X_1-X_2+N=k) = \sum_{x=0}^k \tbinom{N}{x}\tbinom{N}{k-x}\frac{1}{2^{2N}} = \tbinom{2N}{k}\frac{1}{2^{2N}}. 
\end{equation}
 Therefore, $X_1-X_2+N\sim B(2N,1/2)$. Then we have
 \begin{align}
     &\mathbb{P}(|X_1-X_2|\leq N^\alpha) = \mathbb{P}(|X_1-X_2+N-E[X_1-X_2+N]|\leq N^\alpha)\\
    & \leq\sum_{k=-[N^\alpha]-1}^{[N^\alpha]+1} \tbinom{2N}{k}/2^{2N}\leq (2[N^\alpha]+3)\tbinom{2N}{N}/2^{2N}\\
    &\leq  (2[N^\alpha]+3)\frac{2^{2N}}{\sqrt{\pi N}2^{2N}}=\frac{(2[N^\alpha]+3)}{\sqrt{\pi N}}.\label{m=1}
 \end{align}.

  Then we consider m independent variables $X_1, X_2, X_3,\cdots, X_m$. Any two of them satisfy the inequality in Eq.~\eqref{m=1}. Then we have
 \begin{align}
     \mathbb{P}(\min|X_i-X_j|\leq N^\alpha)&= \mathbb{P}(\cup |X_i-X_j|\leq N^\alpha)\\
     &\leq \sum_{i<j}\mathbb{P}(|X_i-X_j|\leq N^\alpha)=\frac{m(m-1)}{2}\mathbb{P}(|X_i-X_j|\leq N^\alpha)\\
     &\leq \frac{m(m-1)}{2}\frac{(2[N^\alpha]+3)}{\sqrt{\pi N}}.
 \end{align}

Then consider the order statistics of $X_i$ is $X_{(1)}\geq X_{(2)}\geq\cdots\geq X_{(m)} $, we have:
\begin{equation}
    \mathbb{P}(|X_{(1)}-X_{(2)}|\leq N^\alpha)\leq \mathbb{P}(\min|X_i-X_j|\leq N^\alpha)
    \leq\frac{m(m-1)}{2}\frac{(2[N^\alpha]+3)}{\sqrt{\pi N}}.
\end{equation}

This concludes the proof of Lemma 1.$\blacksquare$

\section{Projected ensemble of
the tilted-basis model: numerical simulations}
\label{sec:PE_RPS_numerics}

Here we provide details on various aspects of the numerical simulations performed for the tilted-basis model; in particular the estimation of the phase transition point as well as finite-size scaling collapse analysis.

 In what follows, we fix the initial product state  $|\Psi_0\rangle = \left(\cos(\theta_0/2)|0\rangle + e^{i\phi_0}\sin(\theta_0/2))|1\rangle\right)^{\otimes N}$ with   Bloch angles at $(\theta_0$\,$,$\,$\phi_0)$\,$=$\,$(\pi/4$\,$,$\,$\pi/4)$ and then apply a random permutation unitary to reproduce a random permutation state (RPS), before computing the projected ensemble (PE) for a subsystem size $N_A=2$. 
All quantities plotted (e.g., trace distances, or ensemble-averaged coherences) are also averaged over many different instances of RPSs. Importantly, the averaging over RPSs is performed {\it after} computation of the aforementioned quantities.
Unless otherwise stated, for $N = 12,16, 18, 20,22,24$,  the number of samples utilized is by default  $10^4,5\times10^3, 10^3, 5\times10^2, 10^2, 10^2$.
%$10,100,1000,100^3, 10^7??$
% \ww{Chang, please fill in}

% under titled basis $\hat n$ while keeping the subsystem size fixed at $N_A = 2$. 

% Our analytical results for measurements in the $x$- and $z$-bases show that the average form of the PE approaches the Haar ensemble and the classical bit-string ensemble, respectively. However, we expect that for a typical fixed random permutation state  in the tilted-basis model, the corresponding PE will also converge to these two limiting ensembles in the TDL. To verify this, we numerically investigate the PE of the tilted-basis model.

\subsection{Projected ensemble for $x$ and $z$ measurements}

First we provide numerical evidence of convergence of the projected ensemble to the classical bit-string and Haar ensembles, for the special case of $x$- and $z$-measurements respectively, as claimed in the main text and elaborated upon in Sec.~\ref{sec:PE_RPS_analytic}. 

\begin{figure}[t]
\includegraphics[width=0.65\textwidth]{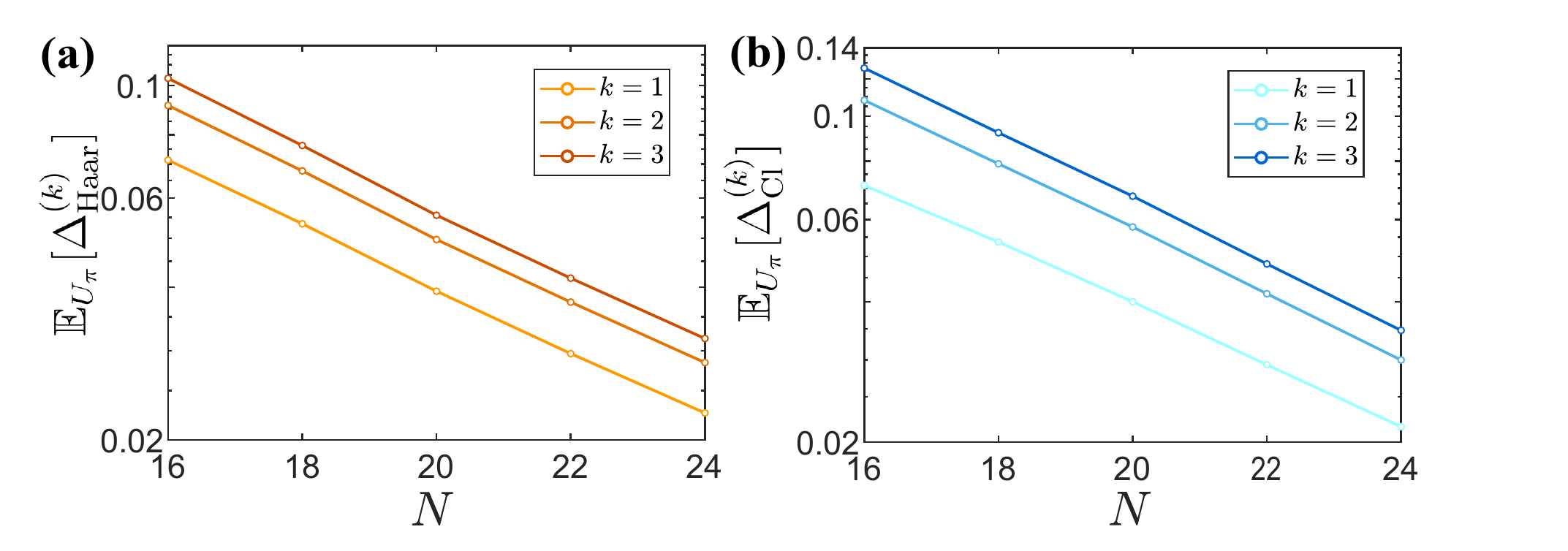}
\caption{  Trace distances of the k-th moments of PE to (a) the  Haar ensemble  and (b) the classical bit-string ensemble (b) versus system size $N$, for $x$- and $z$-basis measurements respectively.
%The PE is constructed from the RPS of the tilted-basis model with initial Bloch angles $(\theta_0$\,$,$\,$\phi_0)$\,$=$\,$(\pi/4$\,$,$\,$\pi/4)$ and fixed $\phi_m = 0$.  (a) and (b) correspond to measurement polar angles $\theta_m = \pi/2$ ($x$-basis) and $\theta_m = 0$ ($z$-basis), respectively.
}
\label{Fig:S1}
\end{figure}

In Fig.~\ref{Fig:S1}, we compute the trace distances $\Delta^{(k)}_{\mathrm{Haar}}$ ($\Delta^{(k)}_{\mathrm{Cl}}$) between the $k$th-moment of the PE under $x$-basis ($z$-basis) and the corresponding $k$-th moment of the Haar (Classical bit-string) ensemble with different system sizes $N$.  Indeed, we find the PE converges exponentially fast to the corresponding limiting ensemble with system size $N$.

\subsection{Estimation of the critical point}

Fig.~1(b) of the main text showed the behavior of the projected ensemble constructed by varying the measurement direction $\hat{n}$ from the $z$- to the $x$-basis,  for its $k=2$ moment.
It indicated a deep thermalization phase transition at $\theta_m^*\approx0.193\pi$ below (above) the limiting behaviors were the classical bit-string (Haar) ensembles.
Here we explain how we determined this crossing point.

To estimate the critical point $\theta_m^*$ and its uncertainty, we utilized the crossing between the two largest system sizes simulatable, $N = 22$ and $N = 24$. We sampled $M = 1000$ and $M = 500$ random permutation realizations respectively, for two nearby measurement angles $\theta_m = 0.19\pi$ and $0.20\pi$ flanking the transition point. From these samples, we obtained the mean trace distance values $\overline{\Delta_{\mathrm{Haar}}^{(2)}}(0.19\pi;N=22)$, $\overline{\Delta_{\mathrm{Haar}}^{(2)}}(0.20\pi;N=22)$, $\overline{\Delta_{\mathrm{Haar}}^{(2)}}(0.19\pi;N=24)$, $\overline{\Delta_{\mathrm{Haar}}^{(2)}}(0.20\pi;N=24)$.
We then employed linear fits to these data from which we extracted the crossing to be $\theta_m^* \approx 0.1930\pi$. The statistical uncertainty was estimated by computing the standard deviation $\sigma(\Delta_{\mathrm{Haar}}^{(2)})$ for each dataset using the unbiased estimator (with denominator $M-1$), followed by computation of the standard error $\delta \Delta_\text{Haar}^{(2)} = \sigma(\Delta_{\mathrm{Haar}}^{(2)})/\sqrt{M}$. Propagating these errors through the linear-crossing relation,
\begin{equation}
(\delta \theta_m^*)^2 = \sum_i \left( \frac{\partial \theta_m^*}{\partial \Delta_{\mathrm{Haar},i}^{(2)}} \right)^2 (\delta \Delta_{\mathrm{Haar},i}^{(2)})^2,
\end{equation}
(here the $i$ runs over the measurement angles and system sizes) 
gives the final estimate with error bars $\theta_m^* \approx 0.1930\pi \pm 0.0002\pi$.
To cross-check, we performed the same analysis using the trace distance $\Delta_{\mathrm{Cl}}^{(2)}$, obtaining $\theta_m^* \approx 0.1931\pi \pm 0.0002\pi$, in excellent agreement with the result from $\Delta_{\mathrm{Haar}}^{(2)}$.
% This consistency confirms that the change in the limiting distribution of the projected ensemble occurs at a well-defined critical point, evidencing a  phase transition.

\subsection{Behavior of other  moments of the projected ensemble for arbitrary measurement direction $\hat{n}$}

\begin{figure}[b]
\includegraphics[width=0.65\textwidth]{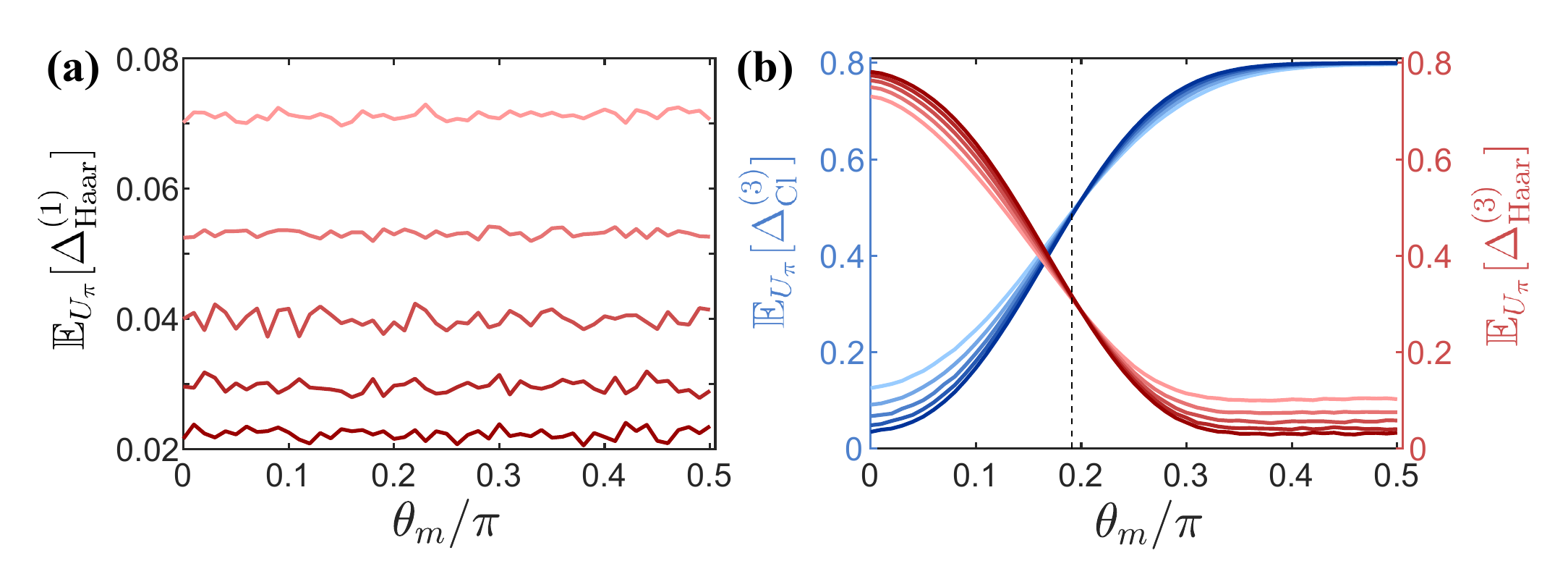}
\caption{  Trace distances of the (a) $k=1$ and (b) $k=3$ moments of projected ensemble to the  Haar ensemble  and the classical bit-string ensemble versus  polar angle $\theta_m$ specifying the measurement basis. Different intensities indicate
different system sizes $ N = 16, 18, 20, 22, 24$ (lighter to darker). For $k=1$ trace distances converge exponentially quickly to zero; for $k=3$ the curves cross at the critical angle $\theta_m^*\approx0.193\pi$, which coincides with the critical point observed for $k=2$ in Fig.~1(b) of the main text.
}
\label{Fig:S2}
\end{figure}
 
We repeat the computation of the trace distances to the touted limiting ensembles for the $k=1$ and $k=3$ moments, to show that (i) there is no transition at the level of the density matrix ($k=1$) [in line with Theorem 1 saying there is only ever infinite-temperature thermalization locally]; and (ii) the transition is also present in the higher moments, with the same crossing point as $k=2$. This is shown in Fig.~\ref{Fig:S2}. 

% Furthermore, as shown in Fig.~2(b) of the main text, the $k=2$ trace distances $\Delta^{(2)}_{\mathrm{Haar}}$ and $\Delta^{(2)}_{\mathrm{Cl}}$ respectively flow toward zero or to their maximal values, demonstrating that the limiting form of the PE in the tilted-basis model undergoes a sharp deep thermalization phase transition as the local measurement direction $\hat{n}$ is varied. Here, we further show the $k=1$ and $k=3$ trace distances for different polar angles $\theta_m$ while fixing $\phi_m = 0$. We find that the $k=1$ trace distances exhibit no crossing, indicating that this phase transition is invisible to the subsystem’s reduced density matrix. By contrast, the crossing point of the $k=3$ trace distances coincides with that of $k=2$, confirming that the two phases correspond to the two limiting ensembles of the PE.

\subsection{Finite-size scaling collapse of the coherence order parameter}

Here we investigate the nature of the critical point by performing finite-size scaling of the order parameter, the ensemble-averaged coherence $\overline{C_r}$, introduced in the main text. 

  \begin{figure}[H]
  \centering
\includegraphics[width=0.65\textwidth]{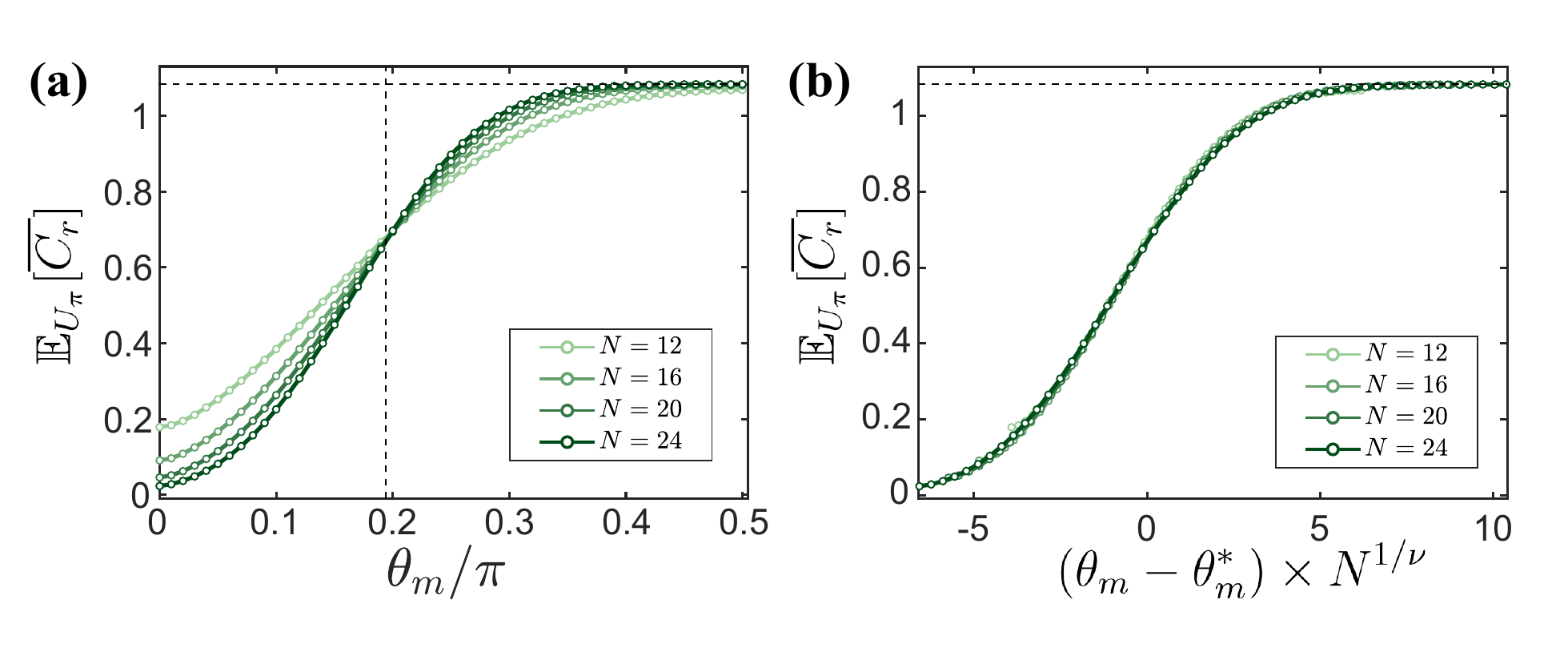}
\caption{ 
(a) Ensemble-averaged coherence $\overline{C_r}$ of the PE constructed from the RPS of the tilted-basis model  with initial Bloch angles $(\theta_0$\,$,$\,$\phi_0)$\,$=$\,$(\pi/4$\,$,$\,$\pi/4)$ and fixed $\phi_m = 0$, showing crossings at $\theta_m^*\approx 0.193\pi$.
(b) Finite-size scaling  collapse of the ensemble-averaged coherence yields $\nu\approx 1.3 \pm 0.2$.}
\label{Fig:S3}
\end{figure}

\begin{figure}[H]
\centering
\includegraphics[width=1\textwidth]{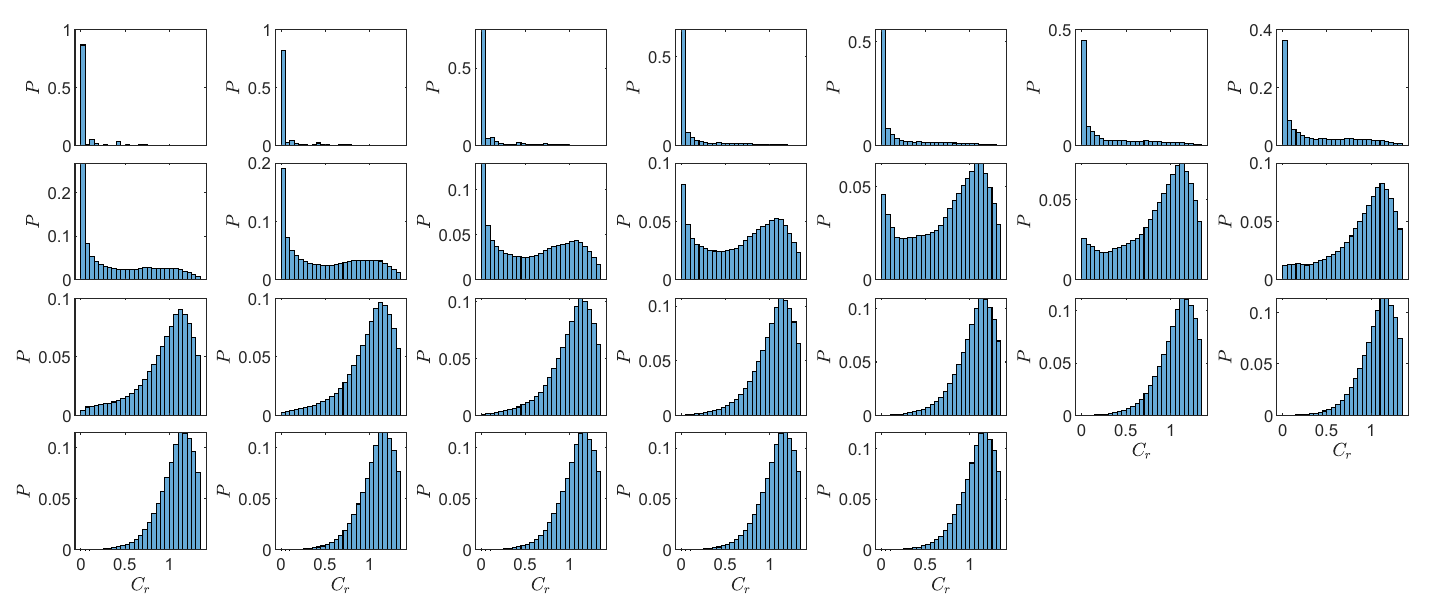}
\caption{ Distributions of the coherence over projected states and permutation realizations in the tilted-basis model for system size $N = 20$. The initial state is prepared with Bloch angles $(\theta_0$\,$,$\,$\phi_0)$\,$=$\,$(\pi/4$\,$,$\,$\pi/4)$, and we fix $\phi_m = 0$. Each panel corresponds to a different measurement angle $\theta_m$, ranging from $0$ to $0.5\pi$ in steps of $0.02\pi$ (arranged from left to right, top to bottom). The critical point $\theta_m^* \approx 0.193\pi$ lies between the 10th and 11th panels.
% \matteo{any chance we can make more efficient use of the space? e.g. 4x7 matrix instead of 5x6 (filling whitespace on the sides) and some identification of the common axes to avoid repeating labels and save whitespace between panels? in Python there are `sharex', `sharey' arguments to do this in subplots, not sure for matlab. Also maybe a numerical value of $\theta_m / \pi$ over each panel?}
% \ww{Chang: could you pretty this up, by any chance? on the other hand I also think it's OK the way it is, for a supplement :P}
% \CL{I use 4x7 and common x axes. I didn't use  common y axes because it wouldn't look good on the second row. I try to add the text of $\theta_m/\pi=\cdots$, but it looks too crowded}
}
\label{Fig:S4}
\end{figure}

% Moreover, as discussed in the main text, this deep thermalization phase transition is driven by the coherence injected by the initial state and the choice of measurement basis. To further investigate this, in Fig~\ref{Fig:S3} we numerically compute the ensemble-averaged coherence $\overline{C_r}$ using the same setup as above. We find that $\overline{C_r}$ flows to zero or to its Haar-ensemble value when $\theta_m$ approaches $0$ or $\pi/2$, respectively, and exhibits a sharp transition at the same critical point $\theta_m \approx 0.193\pi$. 
Fig.~\ref{Fig:S3}(b) show the data collapse of the ensemble-averaged coherence by utilizing the finite-size scaling ansatz $\overline{C_r} =  f((\theta_m-\theta_m^*)N^{1/\nu})$.  We extract the critical exponent by minimizing the variance of the rescaled $\overline{C_r}$  curves to achieve optimal finite-size data collapse, and estimate the error bars by removing a data set corresponding to each system size in turn. 
We see a very good collapse with  a critical exponent $\nu\approx 1.3 \pm 0.2$. 
% \ww{Could you briefly comment on how you extracted critical exponent and the error bar -- can you briefly comment? maybe a sentence or two, like mean-square error? and error bar by removing one data set etc.?  } 

Furthermore, we plot in Fig.~\ref{Fig:S4} the evolution of the distributions $P(\overline{C_r})$ of the order parameter $\overline{C_r}$ as we vary $\theta_m$. As expected, in the limiting values $\theta_m = 0, 0.5\pi$, the distributions are unimodal, and sharply peaked near zero and the value of coherence of a Haar random state respectively. However, near the transition (approximately between the $10$th and $11$th panels), the distribution is broad and continuous. Together with the value of the critical exponent $\nu\approx 1.3 \pm 0.2$, it is suggestive that the transition is of the continuous kind; however, more work needs to be done to pin this down exactly.

%\begin{figure}[h]
%\includegraphics[width=0.85\textwidth]{supp_RPS_disorder.pdf}
%\caption{ random tensor product initial state.need to update N=24 data $\theta_c\approx 0.18\pi?$ }
%\label{Fig:S4}
%\end{figure}

\section{Details of the mixed-basis model}
\label{sec:toy_model}

In this section, we provide further details on the mixed-basis model introduced in the main text, including a rigorous statement on regular thermalization, additional numerical results for alternative choices of initial states, and finite-size scaling data collapse near the critical point.

\subsection{Regular thermalization}
Here we desire to make an analogous statement like Theorem 1, that a typical RPS in the mixed-basis model locally thermalizes. This is equivalent to saying that the generalized random subset states
\begin{equation}
    |\Psi_\pi\rangle = 2^{-\alpha_0 N/2} \sum_{z\in S} (-1)^{f_1(z)}i^{f_2(z)} |z\rangle,
    \label{eqn:subset_state}
\end{equation}
generated by applying a random permutation unitary to the initial state
$|\Psi_0\rangle = |0\rangle^{\otimes (1-\alpha_0)N } |{\mathrm{Y}+}\rangle^{\otimes \alpha_0N}$, are almost always  maximally mixed locally. Here $S$ denotes a random subset determined by the permutation $\pi$, and $f_1$, $f_2$ are Boolean functions which can be treated as pseudorandom for $\alpha_0<1$.

We  follow the same approach as in Sec.~\ref{sec:regular_thermalization}. Specifically, we employ Eq.~\eqref{thermalization_bound} to establish the relevant bounds and compute $\mathbb{E}_{U_\pi}[\text{Tr}[(\rho_A^2)]$). Using the Weingarten calculus together with the permutation-invariant basis states defined in Eq.~\eqref{permutation_basis2}, we obtain
\begin{align}
     \mathbb{E}_{U_\pi}[\text{Tr}[(\rho_A^2)]& = \text{Tr}(S\mathbb{E}_{U_\pi}[\rho_A \otimes \rho_A])=  1/d_A+2^{-\alpha_0N}(1-1/d_A)+1/d(d_A-2+1/d_A)+o(1/d)\\
     &\leq  1/d_A+2^{-\alpha_0N}(1-1/d_A)+1/d(d_A-1+1/d_A),
\end{align}
which leads to the concentration bound
\begin{align}
\mathbb{P}(\|\rho_A -\mathbb{I}_A/d_A\|_1 > \epsilon) & <  \frac{d_A   }{\epsilon^2} \left( \text{Tr} \mathbb{E}_{U_\pi} [\rho_A^2] -  1/d_A \right)\\
&<(2^{-\alpha_0N}(d_A-1)+1/d(d_A^2-d_A+1))/\epsilon^2 \\
&= (d^{-\alpha_0}(d_A-1)+ d^{-1} (d_A^2-d_A+1))/\epsilon^2.
\end{align}
Therefore, as long as $\alpha_0 > 0$, i.e., the initial state has any nonzero density of coherence, regular thermalization hapens in the mixed-basis model with high probability.

% we establish that a sufficient condition for regular thermalization to occur with high probability in the mixed-basis model is to fix $d_A$ (as in our case of interest) and ensure $\alpha_0 > 0$, i.e., that the initial state contains a nonzero amount of coherence. Under these conditions, the reduced density matrix of subsystem~$A$ approaches the maximally mixed state in the thermodynamic limit.

\subsection{Numerical results of the mixed-basis model and finite-size scaling}
\begin{figure}[h]
\includegraphics[width=0.8\textwidth]{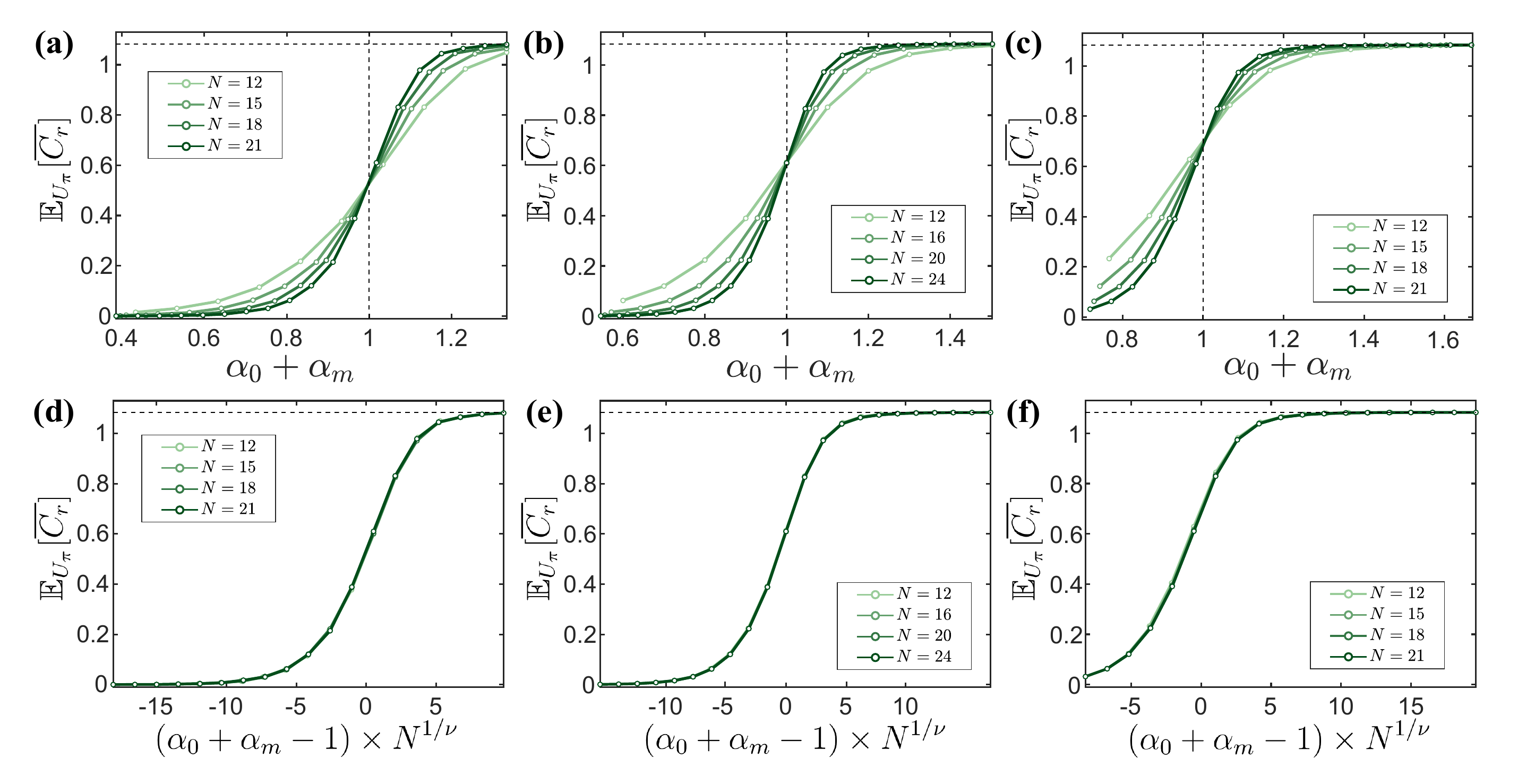}
\caption{  (a-c) Ensemble-averaged coherence of the PE for
mixed-basis model with $\alpha_0 = 0.33\bar{3}, 0.5, 0.66\bar{6}$, showing a sharp transition
at $\alpha_m=0.66\bar{6}, 0.5, 0.33\bar{3}$ as predicted from theory (vertical dashed line).  Horizontal
dashed lines indicate the ensemble-averaged coherence of Haar random states. (d-f) Finite-size scaling collapse of the ensemble-averaged coherence of the PE for
mixed-basis model with $\alpha_0 = 0.33\bar{3}, 0.5, 0.66\bar{6}$ to the form $\overline{C_r} = f((\alpha_0+\alpha_m-1)N^{1/\nu})$. Excellent scaling collapse is seen with $\nu = 1.0\pm 0.2$ in all cases. 
}
\label{Fig:S5}
\end{figure}

In the main text, we presented numerical results for the ensemble-averaged coherence $\overline{C_r}$ of the PE corresponding to the initial state with $\alpha_0 = 0.5$. Here, we extend these results by showing simulations for other values of $\alpha_0$, as displayed in Fig.~\ref{Fig:S5}. We find that the coherence phase transition always occurs at $\alpha_0 + \alpha_m = 1$, in perfect agreement with our analytical predictions in the main text. 

Moreover, we  also perform a finite-size scaling collapse of the coherence curves $\overline{C_r} = f((\alpha_0+\alpha_m-1)N^{1/\nu})$,  demonstrated in Fig.~\ref{Fig:S5}(d-f). By minimizing the variance of the rescaled $\overline{C_r}$ curves to extract the critical exponent, and estimating the error bars by removing one data set for each system size, we find an excellent data collapse with the scaling exponent $\nu = 1.0\pm 0.2$, independent of $\alpha_0$.

% the data collapse of the ensemble-averaged coherence for different initial states with same scaling exponent $\nu\approx0.9$, confirming the universality of the deep thermalization transition across varying $\alpha_0$.

\begin{figure}[H]
\centering
\includegraphics[width=0.8\textwidth]{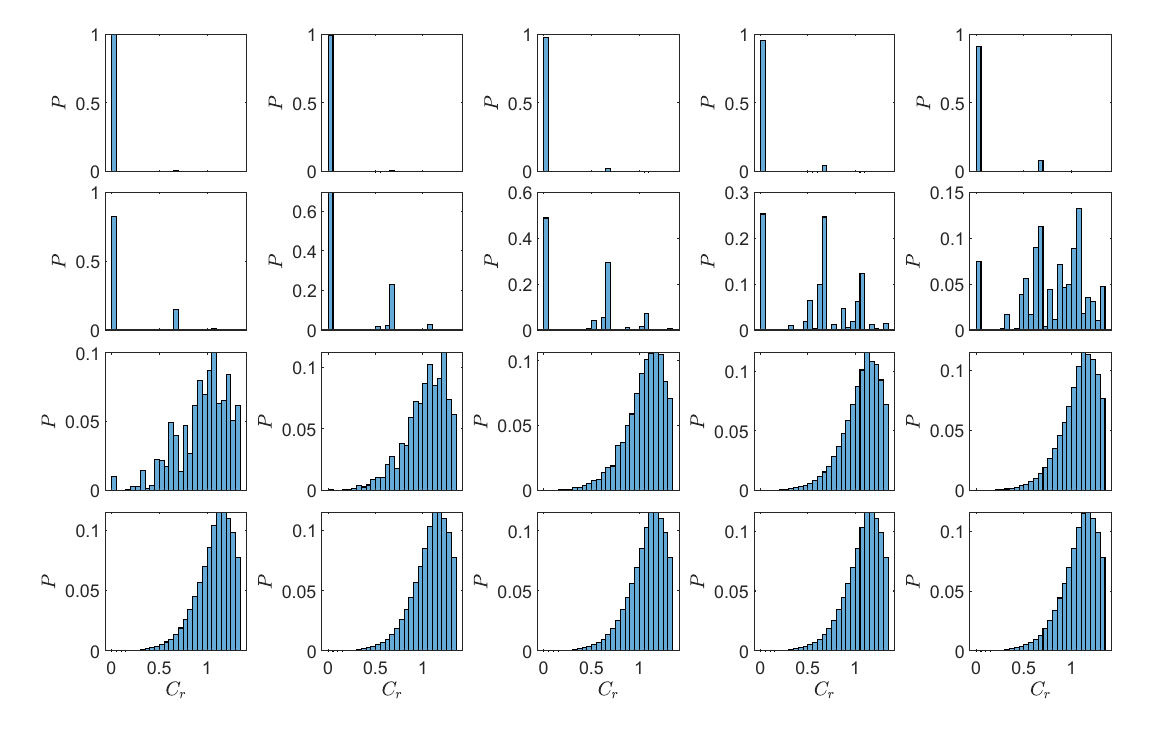}
\caption{ Distributions of the coherence over projected states and permutation realizations in the mixed-basis model for system size $N = 20$ with $\alpha_0 = 0.5$. Each panel corresponds to a different $\alpha_m$, varying from $0.05$ to $1$ in increments of $0.05$ (arranged from left to right, top to bottom). The critical point occurs at $\alpha_m = 0.5$ (10th panel). 
% \matteo{same as for the other plot - here we could make 3x8 instead of 4x6 maybe}\CL{I only use the common x-axis here because I found that the plots in the second row don’t look very good with a common y-axis. So I still use 4x5}
}
\label{Fig:S6}
\end{figure}

%\section{FSS data collapse}
% we can put this part into section PROJECTED ENSEMBLE OF THE RANDOM PERMUTATION STATE.

In \ref{Fig:S6} we also plot the evolution of the distributions $P(\overline{C_r})$ of the order parameter $\overline{C_r}$ as we vary $\alpha_m$ for fixed $\alpha_0 = 0.5$. 
Similar to Fig.~\ref{Fig:S4}, we find that far from the transition $(\alpha_m \approx 0, 1)$, the distributions are sharply peaked near zero and the coherence of a Haar random state respectively. However, unlike Fig.~\ref{Fig:S4}, the distribution at the transition ($10$th panel) does not appear continuous (increasing the number of samples does not change the discrete-looking nature of the distribution). Coupled with the extracted value of the critical exponent $\nu \approx 1.0$ and our   analytical argument of the transition as a simple competition between the low and high coherence phases, it suggests that the transition of the mixed-basis model  is of a first-order kind, in contrast to the tilted-basis model. However, again, more work needs to be done to better pin down the nature of the transition.

\section{Imaginarity of the projected ensemble}
\label{sec:imaginarity} 
In the main text, we analyzed coherence-induced deep thermalization phase transitions in two models, the tilted- and mixed-basis models.
In both cases, the initial states $|\Psi_0\rangle=|\theta_0,\phi_0\rangle^{\otimes N}$ and $|\Psi_0\rangle=|0\rangle^{(1-\alpha_0)N} |\mathrm{Y}_+\rangle^{\alpha_0 N}$ for $\phi_0,\alpha_0 > 0$ were the sole sources of {\it imaginarity} to the system, which is also a resource for quantum information processing \cite{hickey2018quantifying,wu2021resource}. Note in contrast the permutation unitarities implementing dynamics were real; while the measurement bases were also chosen real, as they either lay in the $z$-$x$ plane or were a mixture of $z$- and $x$-measurements.

What would happen if instead we had considered dynamics in which all components were real? Clearly the projected ensemble will also be real regardless of what its limiting form is. The deep thermalized phase thus cannot be the {\it complex} unitary Haar ensemble $\mathcal{E}_\text{Haar}$, characterized by invariance under (complex) unitary rotations.
What then is the appropriate deep thermalized phase?

\subsection{Emergence of real orthogonal Haar ensemble}
We focus on the mixed-basis model for a concrete understanding. For example, we consider an initial state of the form $|\Psi_0\rangle=|0\rangle^{(1-\alpha_0)N} |\mathrm{X}_\pm\rangle^{\alpha_0 N}$, with $|\mathrm{X}_\pm\rangle$\,$=$\,$\frac{1}{\sqrt{2}}(|0\rangle\pm|1\rangle)$.  
Following the theoretical analysis of the  main text of the form of the projected state, it immediately provides the answer: the coefficients of the projected state will no longer be complex Gaussian random variables, but will be real Gaussian random variables.
Therefore, the deep thermalized phase can be  understood to be the {\it real} orthogonal Haar ensemble $\mathcal{E}_\text{O-Haar}$, characterized by invariance under orthogonal rotations (somewhat unsurprisingly). 

\begin{figure}[h]
\includegraphics[width=0.85\textwidth]{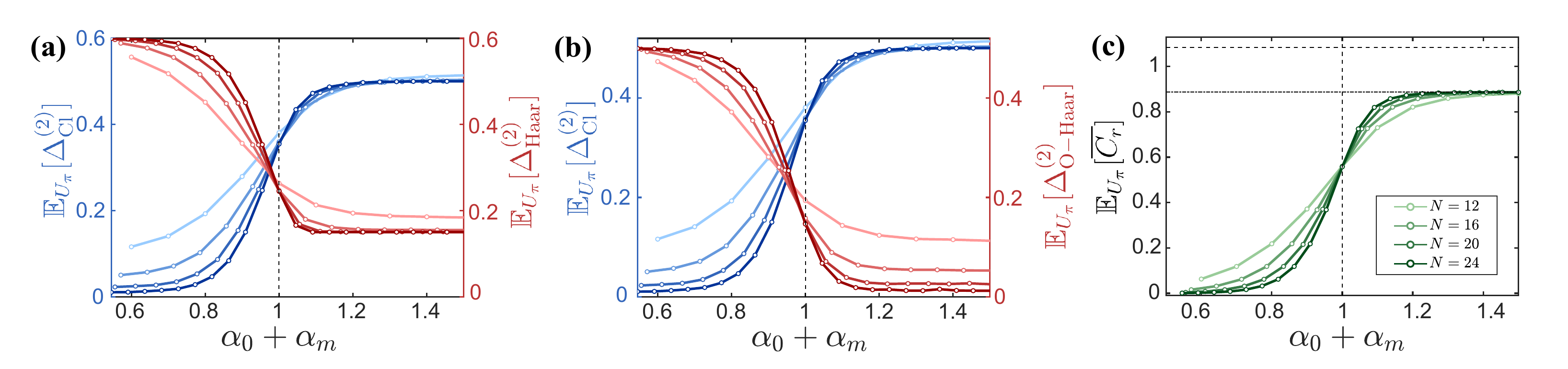}
\caption{(a–c) Trace distances and coherence of the PE with $k = 2$ and subsystem size $N_A = 2$ for the mixed-basis model with initial state $|0\rangle^{\otimes (1-\alpha_0)N} \otimes |\mathrm{X}_-\rangle^{\otimes \alpha_0 N}$ and $\alpha_0 = 0.5$.
(a) Trace distances between the PE and the classical bit-string ensemble as well as the (complex, unitary) Haar ensemble.
(b) Trace distances between the PE and the classical bit-string ensemble as well as the real orthogonal Haar ensemble (``O-Haar''). Color intensity indicates the system size $N = 12, 16,  20, 24$, increasing from lighter to darker shades. The vertical dashed line indicates the transition point at $\alpha_m = 0.5$. The horizontal dashed lines mark the coherence of complex Haar-random states, while the dash-dotted line marks the coherence of real Haar-random states ($\approx 0.886$ for $N_A =2$).
%The vertical dashed line indicates the transition point at $\alpha_m = 0.5$, and horizontal dashed lines mark the coherence of the complex Haar-random states.
% \ww{Chang, could you add one more dashed horizontal line for value of C for real Haar states?
% Can just numerically obtain, don't need closed form expression. perhaps dot dashed or something else}
% \CL{$\overline{C}\approx 0.886$, I run for $10^8$ samples}
}
\label{Fig:S8}
\end{figure}

In Fig.~\ref{Fig:S8}(a-b) we confirm this by performing numerical simulations on the mixed-basis model with $\alpha_0 = 0.5$ and initial state $|\Psi_0\rangle=|0\rangle^{(1-\alpha_0)N} |\mathrm{X}_-\rangle^{\alpha_0 N}$, plotting the trace distances  of the $k=2$ moment of the projected ensemble to that of three ensembles $\mathcal{E}_\text{Cl.}$, $\mathcal{E}_\text{Haar}$ and $\mathcal{E}_\text{O-Haar}$.
As expected, convergence is to the classical bit-string ensemble for $\alpha_0+\alpha_m < 1$ and  to the real orthogonal Haar ensemble for $\alpha_0+\alpha_m > 1$.
Convergence to the complex unitary Haar ensemble is not seen. 
A plot of the ensemble-averaged coherence in Fig.~\ref{Fig:S8}(c) also shows convergence to a value consistent with real Haar random vectors in the deep thermalized phase, lower than that of complex Haar random vectors.

\section{Universality of coherence-induced deep thermalization phase transition  through calculations of the ensemble-averaged inverse participation ratio}
\label{sec:IPR}

% \ww{I think I'm misunderstanding this part of the section --- I think we want to make a case for arbitrary input states and arbitrary measurement bases defined only by their IPRs like in Matteo's original note?  I don't think we want to focus on  the tilted-field model, this weakens the claim of universality} 
% \CL{this calculation is general for arbitrary input states and arbitrary measurement bases, while the parameters $g,h,\lambda\cdots$ in the final expression depend on the $\Psi_0,\nu$. But maybe it is not clear in physics, I rewrote this section and only keep the leading order of the Weingarten function. }

In the main text, we commented that the coherence-induced deep thermalization phase transition in the projected ensemble is a universal phenomenon even going beyond the two microsscopic models analyzed. Here we provide arguments for such a claim, by analyzing the ensemble-averaged inverse participation ratio (IPR) as a proxy for coherence.  Using a non-rigorous approximation, we  compute the IPR of the projected ensemble and argue that its scaling behavior exhibits two distinct phases separated by a sharp transition depending on the IPR of the initial state and measurement basis.

Let $|\Psi_\pi\rangle = U_{\pi}|\Psi_0\rangle$ be an $N$-qubit random permutation state generated from an arbitrary input state $|\Psi_0\rangle$. % We perform a local projective measurement $|\Phi_\nu\rangle\langle\Phi_\nu|_B$ on subsystem $B$, where $|\Phi_\nu\rangle_B$ denote an arbitrary local measurement state  with measurement outcome $\nu$. 
We construct the PE by measuring subsystem $B$ with an arbitrary set of local projective measurements $\{|\Phi_\nu\rangle\langle\Phi_\nu|_B\}$.
We denote the normalized post-measurement state on $A$ by $|\psi_A(\nu)\rangle$ and the corresponding unnormalized state by $|\tilde\psi_A(\nu)\rangle=(\mathbb{I}_A\otimes\langle \Phi_\nu|_B)\,|\Psi_\pi\rangle$. Writing the amplitudes in the computational basis as $\psi_i(\nu)$ and $\tilde\psi_i(\nu)$ ($i\in[ d_A ]$), the IPR is
\begin{equation}
    \mathrm{IPR}(|\psi_A(\nu)\rangle)=\sum_{i} |\psi_i(\nu)|^4
    =\frac{\sum_i |\tilde\psi_i(\nu)|^4}{\big(\sum_i |\tilde\psi_i(\nu)|^2\big)^2}.
\end{equation}
We consider the ensemble-averaged IPR of the PE, also averaging over permutations, $\mathbb{E}_{U_\pi}[\mathrm{IPR}(\mathcal{E}_{\mathrm{PE}})]$. Direct evaluation of this expectation is intractable in closed form, so we adopt the common approximation of replacing the ``quenched averaging'' by ``annealed averaging'', i.e. averaging numerator and denominator separately:
\begin{align}
    \mathbb{E}_{U_\pi}
    [\mathrm{IPR}(\mathcal{E}_{\mathrm{PE}})] &=\mathbb{E}_{U_\pi}\left[\sum_\nu \frac{\sum_i |\tilde{\psi}_i (\nu)|^4}{(\sum_i |\tilde{\psi}_i(\nu)|^2)^2 }\sum_i |\tilde{\psi}_i(\nu)|^2\right]\\
    &\approx 
    \sum_\nu \frac{\mathbb{E}_{U_\pi}[\sum_i |\tilde{\psi}_i (\nu)|^4]}{\mathbb{E}_{U_\pi}[ \sum_i |\tilde{\psi}_i(\nu)|^2 ]},\label{average_IPR}
\end{align}
%where we use the fact that $\sum_i |\tilde{\psi}_i(\nu)|^2$ equals the Born probability $p(\nu)$, and in the final step we kept only the leading contribution $\mathbb{E}_{U_\pi}[p(\nu)]\approx 1/d_B$.  %exact expression is in eq-calss theory section.

These averages can be computed using Weingarten calculus. For the  denominator, we have
\begin{equation}
    \mathbb{E}_{U_\pi} 
    \left[ \sum_i |\tilde{\psi}_i(\nu)|^2\right] = \mathbb{E}_{U_\pi}[ p(\nu) ]\approx1/d_B,
\end{equation}
where we only keep the leading order of the Weingarten function.

The numerator involves a quartic moment:
\begin{align}
\mathbb{E}_{U_\pi}\left[\sum_i |\tilde{\psi}_i (\nu)|^4 \right]&= \mathbb{E}_{U_\pi}[\langle\sigma_{1}|_A(\tilde{\psi}_A(\nu)\otimes\tilde\psi_A^*(\nu))^{\otimes 2}\rangle]\\
&=\sum_{i,j}\text{Wg}^{(4)}(\sigma_i,\sigma_j)(\langle\sigma_{1}|_A\otimes\langle(\Phi_\nu \otimes \Phi_\nu^{*})^{ \otimes 2}|_B) |\sigma_i\rangle\langle \sigma_j|(\Psi_0 \otimes \Psi_0^{*})^{ \otimes 2}\rangle \\
&\approx d^{-1} (\langle\sigma_{1}|_A\otimes\langle (\Phi_\nu \otimes \Phi_\nu^{*})^{ \otimes 2}|_B) |\sigma_1\rangle\langle \sigma_1|(\Psi_0 \otimes \Psi_0^{*})^{ \otimes 2}\rangle \\
   % \mathbb{E}_{U_\pi}[\langle\sigma_{1,A}|(\tilde{\psi}_A(\nu)\otimes\tilde\psi_A^*(\nu))^{\otimes 2}\rangle]
  &= \frac{\mathrm{IPR}(|\Psi_0\rangle) \mathrm{IPR}(|\Phi_\nu\rangle_B)}{d}, \label{unnormalized_IPR}
\end{align}
where $\mathrm{Wg}^{(4)}$ is the fourth-order Weingarten function for the symmetric group, $\sigma_i,\sigma_j$ label the partition basis states defined in Eq.~\eqref{permutation_basis2}, and $|\sigma_{1}\rangle_A=\sum_{a=1}^{d_A}|a,a,a,a\rangle$ is the subsystem-$A$ partition basis state. In the first line we use the vectorization formalism, writing $\mathrm{vec}(|\psi\rangle\langle\psi|^{\otimes 2})=|(\psi\otimes\psi^*)^{\otimes 2}\rangle$ to express moments as inner products with permutation basis vectors. 
In the third line, we retain only the leading  Weingarten term $\mathrm{Wg}^{(4)}(\sigma_1,\sigma_1) = d^{-1}$. The last line follows from $\langle \sigma_1 | (\psi \otimes \psi^*)^{\otimes 2} \rangle = \sum_i |\psi_i(\nu)|^4 = \mathrm{IPR}(|\psi\rangle)$.

Substituting into Eq.~\eqref{average_IPR}, the approximate averaged IPR is
\begin{align}
    \mathbb{E}_{U_\pi}[\mathrm{IPR}(\mathcal{E}_{\mathrm{PE}})] &\approx \frac{d_B}{d_A}\mathrm{IPR}(|\Psi_0\rangle) \mathrm{IPR}(|\Phi_\nu\rangle_B),
\end{align}
Two regimes emerge:
\begin{itemize}
    \item[(i)] $d_B\mathrm{IPR}(|\Psi_0\rangle) \mathrm{IPR}(|\Phi_\nu\rangle_B)\ll d_A$, the ensemble-averaged IPR vanish in the thermodynamic limit, corresponding to the ergodic regime.
    %implying the limiting form of the PE is Haar.  
    \item[(ii)]  $d_B\mathrm{IPR}(|\Psi_0\rangle) \mathrm{IPR}(|\Phi_\nu\rangle_B)\gg d_A$, the IPR diverges, signaling the deeply non-ergodic regime.
    %limiting distribution of the PE is the classical bit-string ensemble.  
\end{itemize}

%we comment on the divergence

We emphasize that the ensemble-averaged IPR (on $A$) is strictly bounded by $1/d_A \le \mathrm{IPR}\le 1$, so any apparent ``divergence'' of the approximate expression is an artifact of the uncontrolled step in which numerator and denominator were averaged separately. This approximation is accurate when fluctuations of the denominator (the Born probability $p(\nu)$) are small relative to its mean; conversely, it fails when $\mathbb{E}_{U_\pi}[p(\nu)^2]\gg 1/d_B^2$. To diagnose this, we compute the second moment of the Born probability using the same Weingarten technique:
\begin{align}
\mathbb{E}_{U_\pi}[p(\nu)^2]&= \mathbb{E}_{U_\pi}[\langle\sigma_{6}|_A|(\tilde{\psi}_A(\nu)\otimes\tilde\psi_A^*(\nu))^{\otimes 2}\rangle]\\
&=\sum_{i,j}\text{Wg}^{(4)}(\sigma_i,\sigma_j)(\langle\sigma_{6}|_A\otimes\langle\Phi_\nu|_B) |\sigma_i\rangle\langle \sigma_j|(\Psi_0 \otimes \Psi_0^{*})^{ \otimes 2}\rangle \\
   % \mathbb{E}_{U_\pi}[\langle\sigma_{1,A}|(\tilde{\psi}_A(\nu)\otimes\tilde\psi_A^*(\nu))^{\otimes 2}\rangle]
  &\approx \frac{\mathrm{IPR}(|\Psi_0\rangle) \mathrm{IPR}(|\Phi_\nu\rangle_B)}{d},
\end{align}
where $|\sigma_{6}\rangle_A=\sum_{a=1}^{d_A}|a,a,b,b\rangle$ is the subsystem-$A$ partition basis state. In the third line, we also retain only the leading-order contribution of the Weingarten function. Under this approximation,  $\mathbb{E}_{U_\pi}[p(\nu)^2]$
coincides with the quartic moment in Eq.~\eqref{unnormalized_IPR}. Consequently, the parameter regime in which the approximate IPR diverges corresponds to the regime where the relative variance of $p(\nu)$ diverges. This explains why the annealed averaging approximation fails in this regime --- its gives an unphysical result. Nevertheless, it {\it does} signal a physical change in behavior of the projected states: 
% \CL{maybe we can put this discussion after  the phase boundary for two specific models}
% \ww{I didn't quite understand what you meant here?}
the different scaling behavior of the ensemble-averaged IPR signals two distinct phases. 

Hence, our calculation can be understood as providing  a  (loose) criterion  $d_B\mathrm{IPR}(|\Psi_0\rangle) \mathrm{IPR}(|\Phi_\nu\rangle_B)\sim d_A$ for the coherence-induced deep thermalization phase transition point. Applying this to our specific models for fixed $d_A$, we   obtain:

%This calculation thus provides a criterion for the deep thermalization phase transition. Applying it to our specific models:

Mixed-basis model:
\begin{equation}
\mathrm{IPR}(|\Psi_0\rangle) = 2^{-\alpha_0 N}, \quad
\mathrm{IPR}(|\Phi_\nu\rangle_B) = 2^{-\alpha_m N_B},
\end{equation}
which reproduces the same phase boundary $\alpha_0 + \alpha_m = 1$ derived in the main text.%recovering the phase boundary $\alpha_0 + \alpha_m = 1$ as in the main text.

Tilted-basis model:
\begin{equation}
\mathrm{IPR}(|\Psi_0\rangle) = (\sin^4(\theta_0/2) + \cos^4(\theta_0/2))^N, \quad
\mathrm{IPR}(|\Phi_\nu\rangle_B) = (\sin^4(\theta_m/2) + \cos^4(\theta_m/2))^{N_B},
\end{equation}
predicting a phase boundary  $\theta_m^{*''} \approx 0.304\pi$ 
for $(\theta_0,\phi_0) = (\pi/4,\pi/4)$, while numerics indicate $\theta_m^* \approx 0.193\pi$. The quantitative discrepancy is expected due to the  ``annealed averaging'' approximation. 

Nevertheless, the virtue of the calculation performed here is that it is generic for arbitrary input states and measurement bases, depending   on only their coherences: 
it demonstrates that the coherence-induced deep thermalization phase transition in the projected ensemble is a \emph{universal} phenomenon.

\section{Coherence-induced deep thermalization phase transition in random permutation dynamics with diagonal phase gates}
\label{sec:monomial}
%Coherence-induced phase transition beyond the permutation dynamics? 
%or generalized permutation unitary?

% \CL{I also added a sentence at the beginning of the appendix}

% In the main text, we studied the coherence-induced phase transition of the projected ensemble under random permutation dynamics (RPD), uncovering that this deep thermalization transition is driven by coherence injected through the initial state and the measurement basis. 
% By the nature of this transition, 
In the main text, we argued that dynamics being coherence-preserving  is a necessary physical constraint for the transition to occur: the transition is driven by whether the global injection of coherence via the initial state and measurement basis, proliferates to a local subsystem or not, under scrambling dynamics which does not otherwise create more superpositions (i.e., is a free incoherent operation).  
For the purposes of our analyses, we had focused on random permutation dynamics, but we can also consider the most general case of unitary evolutions which are free and incoherent.
We show below that this is generated by the set of random permutation unitaries decorated  with diagonal phase gates, i.e., matrices whose entries coincide with some permutation matrix but for which  each nontrivial one can be an arbitrary $U(1)$ phase factor (as opposed to being fixed at unity). 
Further, we present numerical evidence that the same coherence-induced deep thermalization transition we have uncovered is also present in typical instances of random permutation dynamics with diagonal phase gates, bolstering claims of its universality.

%Random permutation unitaries provide a natural example, as they merely relabel computational basis states and therefore preserve the coherence of any initial state. This raises the question of whether random permutation dynamics constitute the most general class of unitaries with this property. In this section, we show that the largest class of unitaries that preserve coherence for arbitrary input states is given by permutation unitaries with diagonal phase gates. We further present numerical evidence that the same phase transition persists under this more general class of dynamics.

To see this, consider a unitary operator $U$ that preserves coherence for arbitrary input states. We first take the input state to be a computational basis state $|z_1\rangle$, which has zero coherence. Since $U$ preserves coherence, the output state must also have zero coherence:
\begin{equation}
    C_r(U|z_1\rangle) = 0.
\end{equation}
Because $U|z_1\rangle$ is a pure incoherent state, it must be a computational basis state up to a phase. Therefore,
\begin{equation}
    U|z_1\rangle = e^{i\phi(z_1)}|z_2\rangle,
\end{equation}
for some computational basis state $|z_2\rangle$ and phase $\phi(z_1) \in \mathbb{R}$. Since $U$ is unitary, this mapping must be bijective over the computational basis, implying that any coherence-preserving $U$ can be written as the application of a diagonal matrix with phase factors followed by a permutation of basis states, i.e., $U = U_\pi D$ for some permutation unitary $U_\pi$ and diagonal matrix $D$. %Hence, the largest set of coherence-preserving unitaries is a subset of the permutation unitaries.

% \textcolor{red}{WW: Chang, I don't  really understand this paragraph, can't we already conclude from (11) and (12) the form $U = U_\pi D$? and by construction it is coherence preserving? So why the need for this calculation? Also, why does the next paragraph start with `Conversely'? What is being contrasted here? What is an apple which cannot be compared to an orange? }
To show sufficiency, we next prove that all permutation unitaries with phase gates preserve coherence of an  arbitrary state $\rho$. 
%Consider a unitary of the form
%\begin{equation}
%    U = U_\pi D,
%\end{equation}
%where $U_\pi$ is a permutation unitary and $D$ is a diagonal unitary. For an arbitrary density matrix $\rho$, 
A simple calculation shows the relative entropy satisfies
\begin{align}
    C_r(U\rho U^\dagger)
    &= C_r(U_\pi D \rho D^\dagger U_\pi^\dagger) \nonumber \\
    &= S\!\left((U_\pi D \rho D^\dagger U_\pi^\dagger)_{\mathrm{diag}}\right) - S(\rho) \nonumber \\
    &= S\!\left((U_\pi \rho U_\pi^\dagger)_{\mathrm{diag}}\right) - S(\rho) \nonumber \\
    &= S\!\left(\sum_z p_z |z\rangle\langle z|\right) - S(\rho) \nonumber \\
    &= C_r(\rho),
\end{align}
where $\{p_z\}$ are the diagonal elements of $\rho$ in the computational basis. In the third equality, we used the fact that diagonal unitaries do not change diagonal elements, and the fact that permutation unitaries merely relabel computational basis states.
Therefore, we conclude that the largest set of coherence-preserving unitaries is precisely the set of permutation unitaries with phase gates.

%------------------------

% Permutation unitaries can be interpreted as classical reversible dynamics, whereas phase gates additionally introduce relative phase imbalances, which are purely quantum effects. Although the microscopic dynamics in these two cases are different, the crucial property relevant to the deep thermalization transition we studied is whether the dynamics preserve coherence.  Therefore, it is natural to expect that the same phase transition persists under random permutation dynamics with phase gates , just as it does under purely random permutations.

% To verify this prediction,

\begin{figure}[t]
\includegraphics[width=0.65\textwidth]{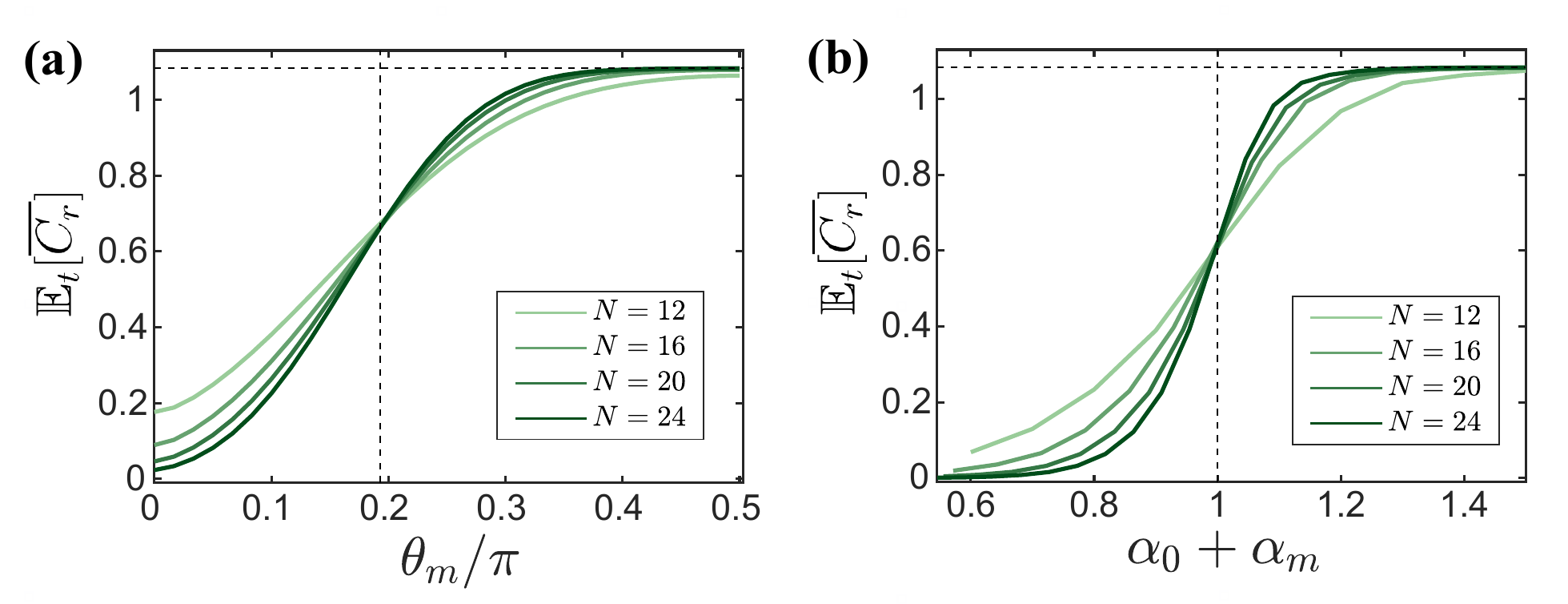}
\caption{ Saturation value of the ensemble-averaged coherence $\overline{C_r}$ of the projected ensemble  for (a) the tilted-basis model with initial Bloch angles $(\theta_0,\phi_0)=(\pi/4,\pi/4)$ and fixed $\phi_m=0$, and (b) the mixed-basis model with $\alpha_0=0.5$. In both cases, the state evolves under the quantum circuit composed of three-local random permutation gates and onsite random phase gates. The coherence saturates to that of the classical bit-string ensemble and the Haar ensemble in the two phases, respectively, demonstrating the same transition as in random permutation dynamics. The critical values (a) $\theta_m^* \approx 0.193\pi$ for the tilted-basis model and (b) $\alpha_m = 0.5$ for the mixed-basis model are indicated by vertical dashed lines. The saturation value is obtained by averaging over the time interval $[2N,\,4N]$.
}
\label{Fig:S9}
\end{figure}

Next, we present numerical results for whether or not random permutation dynamics with diagonal phase gates support a coherence-induced deep thermalization transition. We focus on brickwork circuit models composed of the application of a layer of random three-qubit permutation gates followed by a layer of onsite random phase gates, and study the projected ensemble of states generated by such dynamics in the set-ups of both the tilted-basis and mixed-basis  models. As shown in Fig.~\ref{Fig:S9}, we observe the same transition from the classical bit-string ensemble to the Haar ensemble occurs upon tuning the coherence of either the initial state or the measurement basis, exactly as in the random permutation case of the main text. Moreover, the location of the critical point remains unchanged. This provides further evidence that the coherence-induced phase transition is not restricted to permutation unitaries, but is a generic feature of coherence-preserving dynamics.  We note that  conversely, if one considers a typical unconstrained scrambling unitary, such as a Haar-random unitary, the dynamics generically generate an extensive amount of coherence. In this case, the projected states of local subsystems are also expected to be nearly maximally coherent, and the projected ensemble therefore always appears deeply thermalized independent of the measurement basis or the coherence of the initial state, thereby washing out any sharp transition. This expectation is in fact borne out by a rigorous theorem in~\cite{cotler2023emergent}. 

%Finally, it is interesting to ask whether this phase transition remains robust under dynamics that weakly violate the coherence-preserving condition. We leave this question for future investigation.

\bibliography{biblio}